\documentclass[draft]{agujournal2019}
\usepackage{url} 
\usepackage{lineno}
\usepackage[inline]{trackchanges} 
\usepackage{soul}

\draftfalse

\journalname{JGR: Machine Learning and Computation}

\begin{document}

%
%


\title{Atmospheric chemistry surrogate modeling with sparse identification of nonlinear dynamics}

%
%




\authors{Xiaokai Yang\affil{1}, Lin Guo\affil{1}, Zhonghua Zheng\affil{2}, Nicole Riemer\affil{3}, Christopher W. Tessum\affil{1}}


\affiliation{1}{Department of Civil and Environmental Engineering, University of Illinois Urbana-Champaign, Urbana, IL, USA}
\affiliation{2}{Department of Earth and Environmental Sciences, The University of Manchester, Manchester, United Kingdom}
\affiliation{3}{Department of Atmospheric Sciences, University of Illinois Urbana-Champaign, Urbana, IL, USA}




\correspondingauthor{Christopher W. Tessum}{ctessum@illinois.edu}




\begin{keypoints}
\item We apply the sparse identification of nonlinear dynamics (SINDy) to create a machine-learned surrogate model for an atmospheric chemical mechanism
\item We train the model to learn the chemical dynamics in a compressed latent space with lower computational costs
\item The surrogate model can make stable predictions for all modeled chemical species without noticeable error accumulation
\end{keypoints}

%
%

%
%


\begin{abstract}
Modeling atmospheric chemistry is computationally expensive and limits the widespread use of atmospheric chemical transport models. 
This computational cost arises from solving high-dimensional systems of stiff differential equations. Previous work has demonstrated the promise of machine learning (ML) to accelerate air quality model simulations but has suffered from numerical instability during long-term simulations. 
This may be because previous ML-based efforts have relied on explicit Euler time integration---which is known to be unstable for stiff systems---and have used neural networks which are prone to overfitting. 
We hypothesize that the creation of parsimonious models combined with modern numerical integration techniques can overcome this limitation. 
Using a small-scale photochemical mechanism to explore the potential of these methods, we have created a machine-learned surrogate by (1) reducing dimensionality using singular value decomposition to create an interpretably-compressed low-dimensional latent space, and (2) using Sparse Identification of Nonlinear Dynamics (SINDy) to create a differential-equation-based representation of the underlying chemical dynamics in the compressed latent space with reduced numerical stiffness. 
The root mean square error of the ML model prediction for ozone concentration over nine days is 37.8\% of the root mean concentration across all simulations in our testing dataset. 
The surrogate model is 11$\times$ faster with 12$\times$ fewer integration timesteps compared to the reference model and is numerically stable in all tested simulations. 
Overall, we find that SINDy can be used to create fast, stable, and accurate surrogates of a simple photochemical mechanism. In future work, we will explore the application of this method to more detailed mechanisms and their use in large-scale simulations.

\end{abstract}

\section*{Plain Language Summary}
Atmospheric chemistry modeling is computationally expensive due to complex dynamic processes among numerous chemical species. 
Machine learning techniques have the potential to accelerate these models. 
We compress and group the chemical species to reduce their number and apply a machine-learning algorithm based on sparse regression (SINDy) to create a fast, stable, and accurate surrogate model for a simplified photochemical mechanism. 
The machine-learned surrogate model is 11$\times$ faster than the reference model and is numerically stable in all tested nine-day simulation cases.

\section{Introduction}
Modeling atmospheric chemistry is computationally expensive, which limits its widespread practice and also limits its integration into Earth System Models (ESMs) \cite{brasseur2017AQMtextbook}. 
Typical atmospheric chemical mechanisms include reactions that are highly nonlinear and have characteristic time scales varying across several orders of magnitude \cite{logan1981multiscale}, resulting in a numerically stiff system of ordinary differential equations (ODEs) \cite{verwer1995AQMstiffness}.
The Master Chemical Mechanism (MCM) \cite{jenkin1997mcm1,saunders2003mcm2}—the most explicit of such models—contains 5832 species and 17,224 reactions, whereas models used in large-scale simulations contain more simplified mechanisms where chemical species are lumped by chemical bonds \cite{gery1989cbm4,zaveri1999cbmz} or physical and chemical properties \cite{stockwell1997RACM}.
In all cases above, the resulting system suffers from high dimensionality, meaning it has many state variables that must be repeatedly computed.
It also suffers from high stiffness, meaning that the characteristic time scales of different processes vary by orders of magnitude, requiring the system state to be computed at a large number of intermediate time steps to simulate the time-evolution of the system.
These two phenomena combine to cause the high computational cost of these models.
Here, we will explore methods to reduce the computational cost of an atmospheric chemistry model by reducing its dimensionality and its stiffness.

Previous works have demonstrated the promise of machine learning (ML) for accelerating air quality models. 
Neural networks (NN) have been proven as a universal function approximator \cite{hornik1989nnapproximator} and applied in air quality prediction \cite{gardner1998nnreview}. 
\citeA{boznar1993nn} created a multi-layer NN to make short-term (30 minute) predictions for sulfur dioxide (SO\textsubscript{2}) concentrations around a thermal power plant.
\citeA{viotti2002nnblackbox} trained a general form of a three-layer NN for one-hour air pollutant prediction in an urban area.
\citeA{sousa2007pcann} combined principle component analysis (PCA) with a three-layer NN to make predictions for next-day hourly ozone concentration. 

 Despite the promising results in short-term prediction for air pollutants, challenges remain in long-term recurrent time series prediction, including the compounding accumulation of error \cite{sorjamaa2007longtermtimeseriespred}.
 Increasing modeling capacity—for example by creating deeper neural networks—might increase long-term performance but increases challenges related to overfitting and exploding and vanishing gradients \cite{hochreiter2001nn,srivastava2014dropout}.
An alternative approach relies on combining theory-driven and  and data-driven modeling techniques, which are often considered two distinct fields but can in fact be greatly synergistic and complementary \cite{reichstein2019pinn}. 
A growing field, physics-informed machine learning (PIML), aims to use prior knowledge obtained from observational, empirical, physical, or mathematical insights about the world to improve the performance of an ML algorithm \cite{karniadakis2021piml}. 
Knowledge about the source and nature of the dataset, encoded as an inductive bias \cite{baxter2000model}, can provide a useful constraint on ML models. 
\citeA{keller2019application} created a random forest-based surrogate model for gas-phase chemistry in the GEOS-Chem. They separated long-lived from short-lived species and imposed conservation of atoms for the NO\textsubscript{x} family. Despite the success in short-term prediction, exponential error growth occurred for simulation times longer than a few weeks.
\citeA{kelp2020rnn} used a recurrent training regime to make predictions on the box model CBM-Z \cite{zaveri1999cbmz}, using an autoencoder \cite{kramer1991autoencoder} to reduce the dimensionality from 101 chemical species into 16 features without significant loss in performance and reached two orders of magnitude speedup. However, this model also suffered from exponential error accumulation for simulations longer than a few weeks.

 Variants of NN have been developed to address the limitations with sequential data, such as recurrent neural network (RNN) \cite{elman1990rnn}, long-short term memory (LSTM) \cite{hochreiter1997lstm}, and neural ordinary differential equations (NODE) \cite{chen2018node}; however, they still share some common shortcomings in that they are uninterpretable ``black-box'' models, are highly data-intensive \cite{cabaneros2019review}, need tedious trial-and-error experiments on model structure and size \cite{zhang1998review}, and are unable to extract interpretable information and knowledge from the data \cite{karniadakis2021piml}.
 These variants require a hidden state which generally has a much larger dimension than the input, rendering them compatible with inclusion as an operator in a gridded transport model.
 In addition, training and running a NODE-based model requires iterative numerical integration of the neural network which is prohibitively expensive in large-scale systems \cite{djeumou2022taylor}.  

Alternatives to ``black-box'' methods include parsimonious ``white-box'' modeling to describe experimental data \cite{purnomo2023sparse}, as parsimonious models aim to balance predictive accuracy with model complexity, thus preventing overfitting and promoting interpretability and generalizability \cite{brunton2020fluid}. 
 \citeA{schmidt2009distilling} developed a genetic programming-based approach for detecting conservation laws from experimentally collected data without any prior knowledge. 
 \citeA{quade2016prediction} demonstrated how symbolic regression can be applied to identify the future state of dynamic systems.
Finally, \citeA{brunton2016sindy} developed a sparse regression-based method to identify a nonlinear dynamic system, which is the approach we will use here owing to its successful previous application to a wide range of problems \cite{lai2019sparse,hoffmann2019reactive,wang2021inference,jiang2021modeling,pasquato2022sparse}.

\begin{figure}[h]
\centering
 \includegraphics[width=0.8\textwidth]{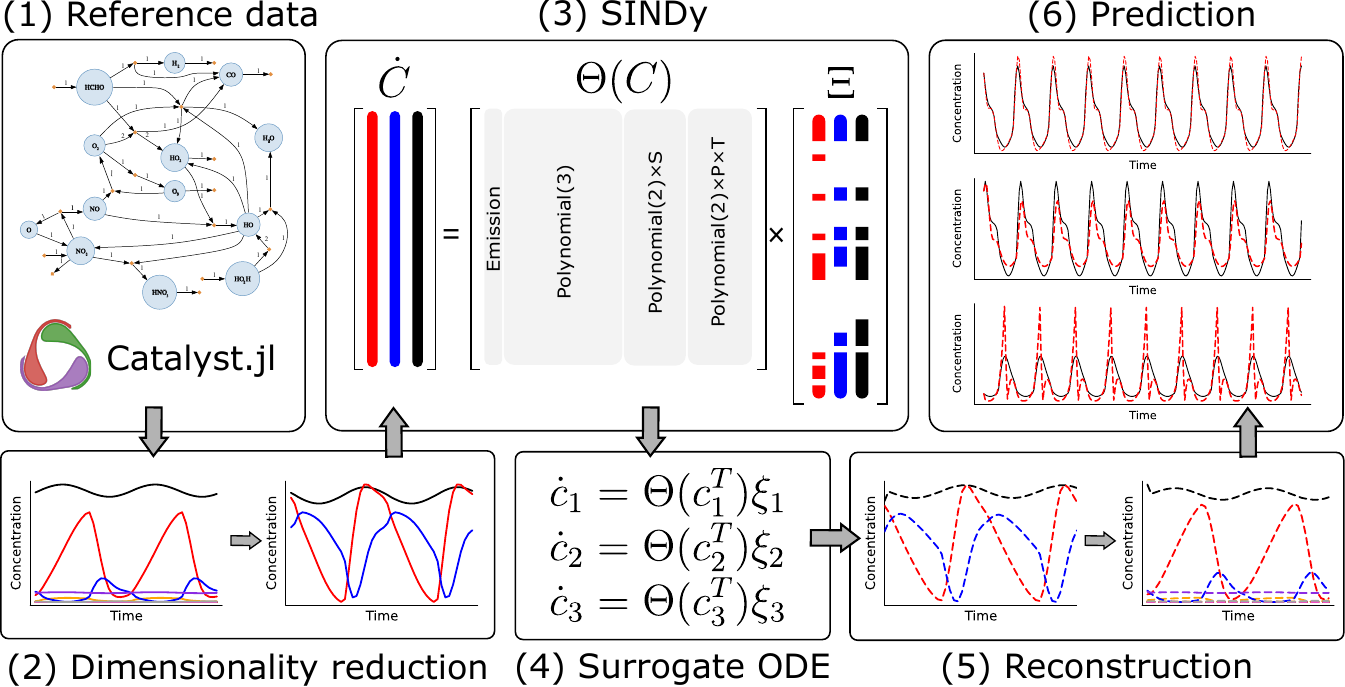}
 \caption{Overall schematic for creating SINDy-based ML surrogate models of an atmospheric chemical mechanism: (1) Run reference photochemical mechanism to generate reference data (Section \ref{Dataset Preparation}); (2) Reduce the dimensionality by compressing data into a latent space to reduce the computational cost of the surrogate model (Section \ref{Dimensionality Reduction}); (3-4) Create a surrogate model based on the Sparse Identification of Nonlinear Dynamics (SINDy) (Sections \ref{SINDy} and \ref{Training Procedure}); (5) Transform data from latent variables back to original variables; (6) Test performance (Section \ref{Surrogate Model Accuracy}).}
 \label{fig:fig_OverallSchem} 
\end{figure}

In this work, we create a machine-learned surrogate model to make fast, stable, and accurate predictions of air pollutant concentrations without the exponential error growth observed in previous work.
We achieve this by (1) reducing the dimensionality through singular value decomposition to create a compressed low-dimensional latent space consisting of latent species with high interpretability and (2) using sparse identification of nonlinear dynamics (SINDy) to train a model to represent the underlying chemical dynamics in the compressed latent space (Figure~\ref{fig:fig_OverallSchem}). To our knowledge, previous works have either reported numerical instabilities or have not included experiments that quantify numerical stability, and we are the first to report numerical stability over long simulation periods under a wide range of conditions.

\section{Materials and Methods}

\subsection{Dataset Preparation} \label{Dataset Preparation}
\begin{figure}[h]
 \centering
 \includegraphics[width=0.5\textwidth]{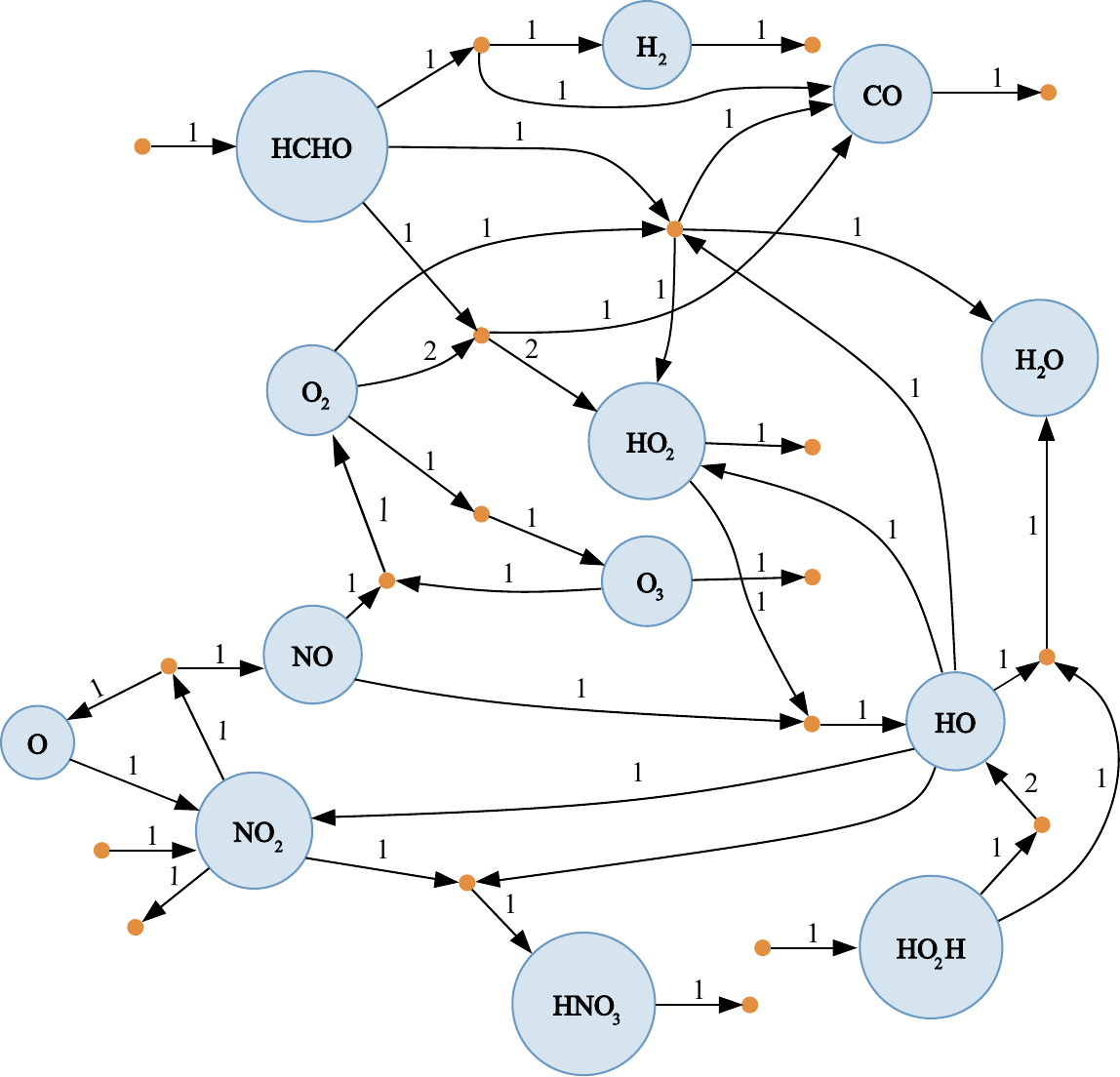}
 \caption{Reference photochemical mechanism used for surrogate training. Black arrows from species to reactions (brown dots) indicate reactants, and are labeled with their input stoichiometry. Black arrows from reactions to species indicate products, and are labeled with their output stoichiometry. Reactions are also shown in Table~S1.}
 \label{fig:fig_RefMech}
 \end{figure}

We aim to emulate a simple photochemical mechanism \cite{sturm2020mb,sturm2022conservation} which contains 11 chemical species and 10 essential reactions related to ozone (O\textsubscript{3}) formation: nitrogen oxides (NO\textsubscript{x}) chemistry, volatile organic compound (VOC) chemistry, formation of a peroxy radical from VOC chemistry, and further reaction of the peroxy radical with nitrogen monoxide (NO) to form nitrogen dioxide (NO\textsubscript{2}) and hydroxyl radical (OH) (Figure~\ref{fig:fig_RefMech}; Table~S1). 

We use Sobol sampling to generate 3750 sets of random initial conditions including temperature, pressure, emission strength, and time phase for time-varying emissions, and a starting time for calculating the evolution of the solar zenith angle. We randomly sample the initial concentrations of the chemical species following \citeA{sturm2020mb}, as shown in Table~S2.
For each emitted chemical species, the emission flux is set as the sum of two components: the first component is a constant background emission, and the second component is a sine function with a period of 24 hours to give the emissions a diurnal pattern (Equation \ref{eq:emis}):

\begin{linenomath*}
  \begin{equation} \label{eq:emis}
   E(i) = 0.95 \times e(i) + 0.05 \times e(i)  \sin(\frac{\pi}{12}t - t_0(i))
 \end{equation}
\end{linenomath*}
where $E(i)$ is the total emission strength of species $i$, $e(i)$ the randomly-sampled emission strength of species $i$, $t_0(i)$ is the starting time (or phase shift) of the emission of species $i$, and $t$ is time in hours.
To prevent the chemical species from accumulating in the model box, we remove O\textsubscript{3}, HNO\textsubscript{3}, HO\textsubscript{2}, NO\textsubscript{2}, CO, and H\textsubscript{2} from the model box at a constant rate (Table~S2).
Consistent with our goal of qualitatively representing atmospheric chemical dynamics so that we can study methods for surrogate modeling, we choose the removal rates empirically to prevent accumulation, rather than measuring or theoretically deriving them.

We partition the 3750 sets of initial conditions into training, validation, and testing sets with a ratio of 8:1:1, using 3000 of the randomly sampled initial conditions to conduct simulations over 72 hours to create a training dataset of 3000 cases and using the remaining two sets of initial conditions (each containing 375 cases) to conduct simulations over 240 hours to create a validation and a testing dataset. (To test generality and numerical stability, our test simulations are for longer periods than our training simulations.) Solutions are saved every 60 minutes of simulation to create our training and testing time series.
To avoid initial transient conditions and unrealistic system behavior \cite{jacobson1997effect,glynn2005initial}, we discard the first 24-hours of data from the training, validation, and testing datasets. 

\subsection{Dimensionality Reduction}\label{Dimensionality Reduction}
As discussed in the introduction, a main cause of the computational intensity of an atmospheric chemistry model is its high dimensionality---the large number of state variables that must be computed at each time step.
Therefore, one way to reduce the computational cost of our model is to reduce its dimensionality.
There is also a secondary benefit of dimensionality reduction, which is that lower-dimensional systems are easier to fit machine-learned models to \cite{bellman1966dynamic, ayesha2020overview, vong2019additional}.

Here we use an explicit and automatic approach based on principal component analysis (PCA) \cite{jolliffe2016principal} to reduce the dataset dimensionality. 
We standardize the dataset by subtracting its mean value over the simulation period across all cases and perform singular value decomposition on the standardized data. 
We calculate the explained variance ratio \cite{MACKIEWICZ1993303} which quantifies how much of the total variance in a dataset is accounted for by each singular value decomposition (SVD) mode. In our case, each SVD mode is a linear combination of chemical species, so we refer to our SVD modes as ``latent chemical species'', which are groups of chemical species lumped automatically during the dimensionality reduction.
Each latent species represents a combination of the original species that is uncorrelated with all other latent species, with each latent species sequentially representing as much variance within the training dataset as possible.
Therefore, to reduce the dimensionality of our model, we can choose a subset of the latent chemical species that explains a sufficient fraction of the overall variance in the training dataset and discard the remaining latent species.

We select the number of the dimensions to which the dataset is compressed, take the same number of columns, or singular vectors, from the left singular matrix, and multiply the dataset array by the selected singular vectors. 
The resulting product array is the representation of the system state in our latent space, where each dimension records the evolution of a latent species that is a linear combination of original chemical species. 
To reconstruct the original species concentrations from the latent representation, we conduct the reverse transformation by multiplying the latent space array with the transpose of the same singular vectors. 
Reconstruction loss inevitably occurs during such reverse transformation, but it can be negligible if the latent species explain a sufficiently high fraction of the overall variance in the dataset.

For public health and regulatory purposes, a specific group of air pollutants including ground level ozone is most important \cite{weschler2006ozone}, while other species such as the intermediate radicals are largely included in models to help to improve the prediction of the important pollutants \cite{kelp2020rnn}. 
Therefore, to ``focus'' our model on predictions of ozone, we increase the importance of ozone in our latent species by multiplying its concentration by a predetermined coefficient ($\beta$) before singular value decomposition to artificially enlarge the fraction of overall variance explained by ozone. 
After compressing the original dataset into a latent space, latent species containing ozone will be prioritized by a factor of $\beta$. 
To reconstruct the ozone concentration, we divide it by the coefficient $\beta$ after the reverse transformation.
The optimal value of $\beta$ is determined through hyperparameter optimization as described in Section \ref{Training Procedure}.

\subsection{Stiffness Reduction}\label{Stiffness Reduction}
 The definition of ODE stiffness is ambiguous but can be characterized by dynamics with widely separated time scales \cite{kim2021stiff}. 
 In atmospheric chemical mechanisms, it means some chemical species evolve much more rapidly---or some chemical processes are much faster---than others. 
 Adaptive timestep numerical methods have been developed to capture both fast and slow dynamics, but choosing proper timesteps is computationally expensive as it includes heavy computational tasks such as calculating the Jacobian matrix \cite{huang2022neural}.
 Here we encourage our surrogate model to be less stiff than the original by training it on datasets of one-hour timesteps. Such selection of time step prioritizes the model's identification of relatively slow, extended chemical processes by smoothing out the fast, instantaneous ones related to species of short lifetimes (such as intermediate radicals). 
 Prioritizing non-stiff dynamics during training results in surrogate models that are less computationally expensive.

\subsection{Sparse Identification of Nonlinear Dynamics}\label{SINDy}
We create a surrogate model based on the Sparse Identification of Nonlinear Dynamics (SINDy) \cite{brunton2016sindy}. 
SINDy is a data-driven method for identifying the governing equations of a dynamical system based on observations of the system's dynamics. 
It is based on the assumption that the dynamics of a given system can be represented as a linear combination of a small number of candidate function terms, each multiplied by a coefficient that represents the strength of that term in the dynamics. 
In the present case, the kinetic dynamics of an atmospheric chemical mechanism can be expressed in the form of a set of ordinary differential equations (ODEs):
  \begin{linenomath*}
  \begin{equation}
\dot{c_i} = \frac{d}{dt}c_i(t) = f(c_i(t))
 \end{equation}
  \end{linenomath*}
where $\dot{c_i}$  is the time derivative of the concentration of chemical species $i$ (its net reaction rate) and $f(c_i(t))$ is the right-hand side (RHS) of the ODE which is defined by the law of mass action.
We assume the underlying dynamics in the ODE can be represented by:
  \begin{linenomath*}
  \begin{equation}
\dot{C} = \Theta (C)\Xi
 \end{equation}
  \end{linenomath*}
where $C = [c_1, c_2, \dots, c_n]$ is the vector of the states, $\dot{C}$ is the vector of time derivatives of the states, $\Theta(C) = [\theta_1(C), \theta_2(C), \dots \theta_n(C)]$ is composed of column vectors of candidate function terms that can define the dynamic system, and $\Xi = [\xi_1, \xi_2, \dots \xi_n]$ is a sparse matrix composed of column vectors of coefficients representing the strength of the active candidate function terms.

A sparsifying algorithm is used to solve the regression problem $\dot{C} = \Theta\left(C\right)\Xi$, where $\dot{C}$ is a matrix of training data, to obtain the sparse coefficient matrix $\Xi$. 
In this problem, each row of $\dot{C}$ is an observation of a rate of change and each column is a state variable of the system of interest.
Each row of $\Theta\left(C\right)$ corresponds to observation in $\dot{C}$ and each column represents a candidate function term.
Each row of $\Xi$ corresponds to a candidate function term and each column represents a state variable. 
(Figure~\ref{fig:fig_OverallSchem} contains a graphical overview.)
In SINDy, sequentially thresholded least squares iteration (STLSQ) \cite{brunton2016sindy,datadrivendiffeq} is implemented by iteratively solving a sparse regression problem on the training data and the candidate function library, and coefficients less than a chosen threshold are set to zero, resulting in a sparse coefficient matrix $\Xi$. 
A common goal of SINDy is to discover equations governing a data-generating process.
Here, we instead use it to discover a system of equations that is less computationally expensive than the reference ODEs we use to generate the training data.

\subsection{Training Procedure}\label{Training Procedure}
We use SINDy to train the surrogate model to emulate the latent variables compressed from the reference model. 
We define and posit a library of candidate function terms that could appear on the right-hand-side (RHS) of the governing equations, including chemical species emission rates, removal rates, polynomials of latent species concentrations, solar zenith angle, pressure, and temperature (Table~S3). 
We use the sequentially thresholded least squares method implemented in the Julia \cite{julia} library DataDrivenDiffEq.jl \cite{datadrivendiffeq}. 

SINDy attempts to find a model that explains rates of change in the training dataset, and does not guarantee that the resulting model will be numerically stable when used in a time-series simulation. 
In practice, we find that our SINDy models do sometimes become numerically unstable, but we are able to remedy this by adding a buffer term---a higher-order polynomial term with a small, negative weight coefficient ($\epsilon$)---to each equation in our SINDy-discovered model as suggested by \citeA{hirsh2022sparsifying}. 
If the library of candidate functions includes polynomial terms up to order $n$, we add a term $- \epsilon x^{n+1}_i$ if $n$ is even, or $-\epsilon x^{n+2}_i$ if $n$ is odd (Figure~\ref{fig:fig_ModelEqn}).
Moreover, we construct our model in a manner that allows us to use modern numerical integration methods (Section \ref{Computational Speed}) rather than the Euler explicit time stepping which we hypothesize has contributed to the numerical instability observed in previous work \cite{kelp2020rnn}.

To improve the model performance and generalizability, we optimize the hyperparameters using random search \cite{bergstra2012random} in Hyperopt.jl \cite{bagge2018hyperopt}.
Hyperparameters include the threshold of the least square iteration ($\lambda$), the weight coefficient of ozone ($\beta$), and the weight coefficient of the buffer term ($\epsilon$) (Table~\ref{tab:1}). We set the optimization metric as the root mean square error (RMSE) between the true value and model prediction on the validation dataset. 
After finishing hyperparameter optimization, we repeat the whole training procedure using the selected set of hyperparameters and evaluate the model performance on the separate testing dataset.
\begin{table}
\caption{Sampling space of hyperparameter optimization}
\label{tab:1}
\centering
\begin{tabular}{cccc}
\hline
Hyperparameter &  Sampling    & Lower bound & Upper bound \\
\hline
$\lambda$      & log-uniform & $10^{-6}$     &$10^{-4}$ \\
$\epsilon$     & log-uniform & $10^{-6}$     &$10^{-4}$ \\
$\beta$        & uniform     & $0.5$         &$10.0$    \\
\hline
\end{tabular}
\end{table}

\subsection{Computational Speed}\label{Computational Speed}
We compare the speed of the surrogate model and the reference photochemical mechanism for simulating 3000 cases on randomly sampled initial conditions over 48 hours. We calculate the total simulation time and the time required per integration time step for each model.

We create a symbolic reaction network system for the reference mechanism using the Julia package Catalyst.jl \cite{2022Catalyst} and ModelingToolkit.jl  \cite{ma2021modelingtoolkit}. 
After training the surrogate model, we rebuild it using the ModelingToolkit.jl \cite{ma2021modelingtoolkit}. We convert the reference model into an ODE system and solve them using the Rosenbrock23 (order 2/3 L-Stable Rosenbrock-W, an ODE solver for stiff problems) solver, and convert the surrogate models into ODE systems and solve them using Tsit5 (Tsitouras 5/4 Runge-Kutta method, an ODE solver for non-stiff problems) solver in DifferentialEquations.jl \cite{shampine1997matlab, tsitouras2011runge, rackauckas2017differentialequations,rackauckas2019confederated}. 
We measure and evaluate the model performance using BenchmarkTools.jl \cite{BenchmarkTools2016}. 
We run all simulations on a single thread of an Intel Xeon Gold 6248 CPU core.

\section{Results}
\subsection{Latent Space Interpretation}
\begin{figure}[h]
 \centering
 \includegraphics[width=0.9\textwidth]{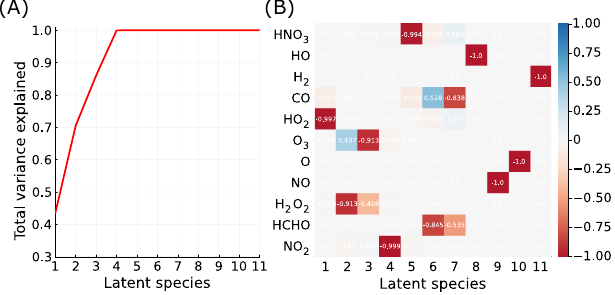}
 \caption{Latent space variance and species mapping. (A) shows the total variance explained by latent species, and (B) shows the mapping relations between chemical species and latent species.}
 \label{fig:fig_LatentSpace}
\end{figure}

Here we explore the interpretability of our latent chemical species. 
The original training dataset has a dimension of 11 chemical species. 
We calculate the fraction of total variance explained as the function of the number of latent species (Figure~\ref{fig:fig_LatentSpace}A). 
Starting from one latent species, the fraction of total variance increases as additional latent species are added and reaches around 100\% at four latent species. 
Latent species after the fourth one can be considered redundant as the fraction of total variance explained by those species is near zero. 
From the perspective of keeping the number of variables low without losing chemical details, we consider three latent species to be sufficient to represent the dataset generated from the reference chemical system, as three principal components can represent over 85\% of the total variance. 

In the reference chemical mechanism, HO\textsubscript{2}, H\textsubscript{2}O\textsubscript{2}, O\textsubscript{3}, and NO\textsubscript{2} correspond to the four most important latent species, as they outweigh the other chemical species in the singular vectors (Figure~\ref{fig:fig_LatentSpace}B). CO and HCHO are allocated across latent species~6 and~7, while HO, NO, O, and H\textsubscript{2} are mapped to latent species~8,~9,~10, and~11, respectively. 
 In the mechanism, O\textsubscript{3} is the species we give extra weight to, and NO\textsubscript{2} is an emitted species and plays a significant role in both the production and destruction of O\textsubscript{3}. 
H\textsubscript{2}O\textsubscript{2} and HCHO are also emitted and products of their decomposition reaction participate in the interrelated processes of ozone formation.
Species such as CO, HNO\textsubscript{3}, and H\textsubscript{2} have low importance because they do not react once being produced (they just accumulate in the model box) and they have a high correlation with those important reactant species. 
Therefore, as four latent species account for almost 100\% of the total variance explained, our latent space representation suggests that HO\textsubscript{2}, H\textsubscript{2}O\textsubscript{2}, O\textsubscript{3}, and NO\textsubscript{2} can represent the reference chemical system while other species such as H\textsubscript{2} can be treated as redundant for the purpose of explaining the variance in the system.

\subsection{Model Interpretation}
\begin{figure}[h]
\centering
 \includegraphics[width=0.75\textwidth]{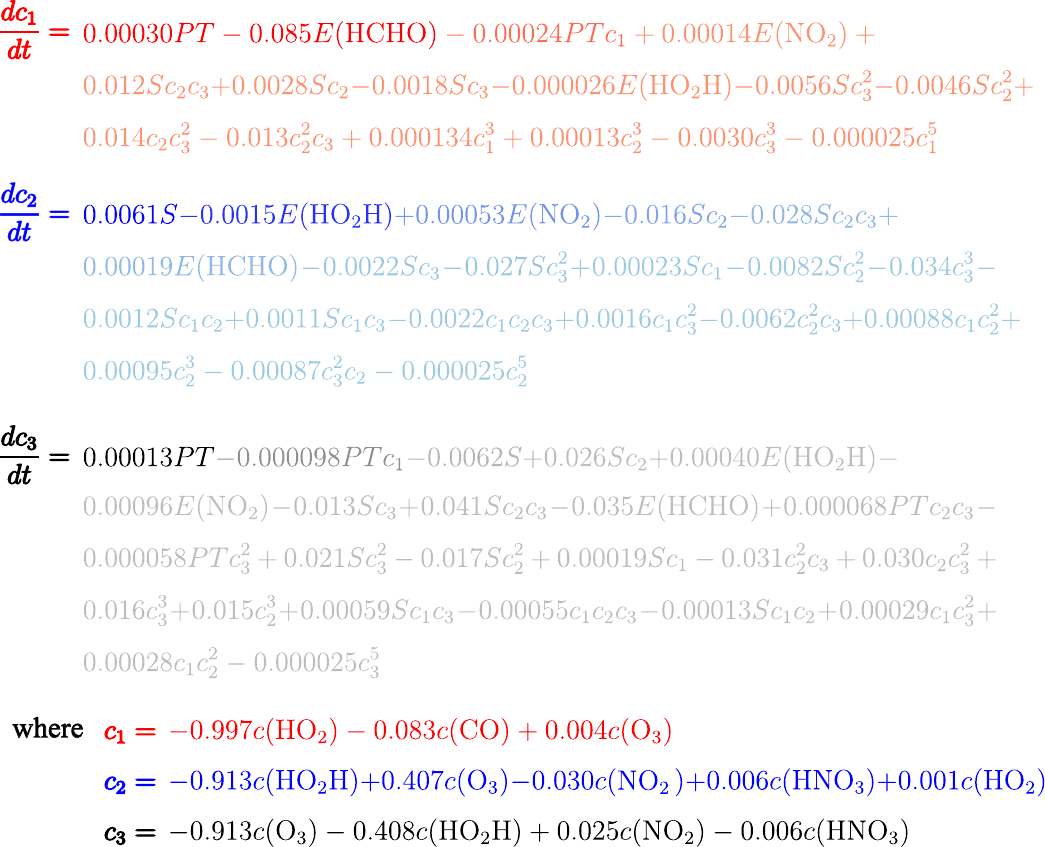}
 \caption{Surrogate model equations. Each term on the RHS is ordered and colored according to the fraction of variance it explains in the training dataset (Table~S4) for each latent species. $c_{i}$ is the concentration of latent species $i$, $c(name)$ is the concentration of chemical species $name$, $P$ is pressure in atm, $T$ is temperature in K, and $S$ is solar zenith angle. }
 \label{fig:fig_ModelEqn}
 \end{figure}

Following the above-mentioned training procedure, we obtain interpretable surrogate ODEs with sparse representation (Figure~\ref{fig:fig_ModelEqn}). 
The concentration of the first latent species ($c_1$) is approximately reversely mapped to HO\textsubscript{2}. The product of pressure and temperature, emission of HCHO account for 46.6\%, 41.5\% of the total variance in the time derivative of $c_1$, respectively. 
The second latent species ($c_2$) can be considered as a linear combination of H\textsubscript{2}O\textsubscript{2} and O\textsubscript{3}. The solar zenith angle, emission of H\textsubscript{2}O\textsubscript{2}, and emission of NO\textsubscript{2} account the most of the total variance (40.2\%, 29.8\%, and 10.5\%) in its derivative. 
Similarly, the third latent species ($c_3$) is dominated by O\textsubscript{3} and H\textsubscript{2}O\textsubscript{2}, and its time derivative is highly dependent on the product of pressure and temperature (73.1\%) and the product of pressure, temperature, and the concentration of the first latent species (17.7\%). 
 
\subsection{Surrogate Model Accuracy} \label{Surrogate Model Accuracy}
In this work, we focus on creating a chemical mechanism surrogate model that can make accurate predictions over a nine-day long period without exponential error accumulation. 
Here we choose to focus on predicting ozone concentration, given its importance in relation to public health and regulation and the limited set of chemical species in our reference model, but the methods here could be used to prioritize a different pollutant or set of pollutants just as easily. 
We optimize the ozone weight coefficient ($\beta$) by random search as described in Section \ref{Training Procedure}. 

\begin{figure}[h]
 \centering
 \includegraphics[width=\textwidth]{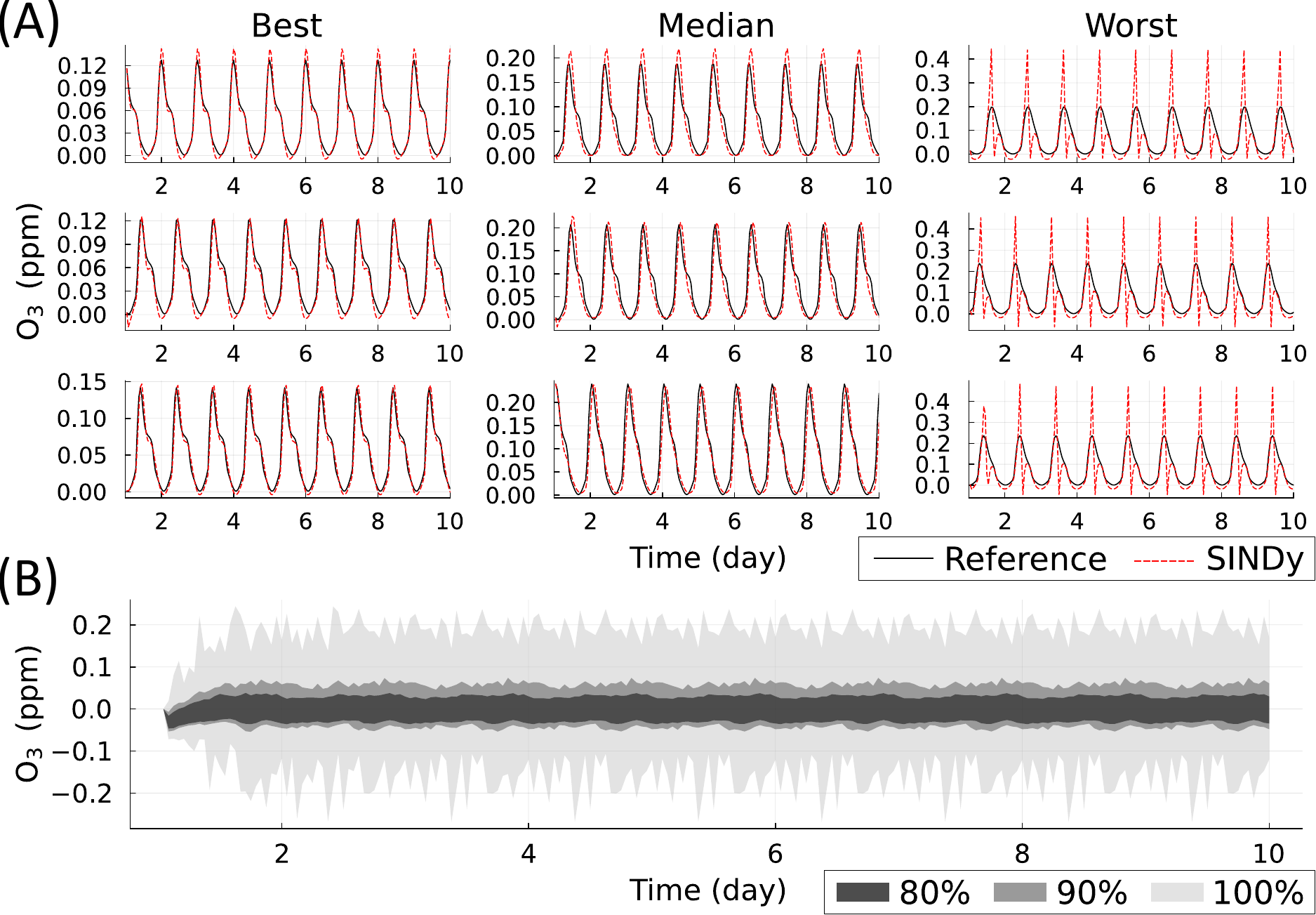}
 \caption{Surrogate model performance. (A) Reference (black solid line) and the surrogate (red dashed line) trajectories for three each of cases with lowest (``best''), median, and highest (``worst'') RSME respectively. (B) Absolute error percentiles for surrogate model testing simulations where the shaded area is the fraction of simulations that has a lower absolute error than the value in the legend. }
 \label{fig:fig_ModelPerformance}
\end{figure}

The overall RMSE of the ML model prediction on the testing dataset for O\textsubscript{3} concentration over nine days is 0.034~ppm, as compared to a 0.090~ppm root mean O\textsubscript{3} concentration across the simulations. We then sort the 375 testing cases by RMSE between the reference model prediction and the surrogate model prediction for O\textsubscript{3} and visualize the three each of the best, median, and worst cases as representatives to illustrate model performance (Figure~\ref{fig:fig_ModelPerformance}A). The model also makes simultaneous predictions for the other species (Figure~S1 to S10).
The model prediction trajectories for the best three simulations follow the diurnal cycle correctly and have RMSEs for O\textsubscript{3} of 0.0078, 0.0095, and 0.0099 ppm, which are 8.67\%, 10.5\%, and 11.0\% of the RMS O\textsubscript{3} concentration (0.090~ppm) in the testing dataset. 
Similarly, the RMSEs in the three median cases are all 0.026 ppm. 
In the worst testing cases, the trajectories are periodic, but amplitudes briefly spike to values much larger than the true function. 
Although the highest RMSE is 0.075~ppm, the trajectories of the worst cases do not suffer from exploding gradients and the error does not accumulate during the simulation period (Figure~\ref{fig:fig_ModelPerformance}B), which represents an improvement over previously published results \cite{kelp2018orders,kelp2020rnn}.
 
Considering the error percentiles for the testing cases (Figure~\ref{fig:fig_ModelPerformance}B), most testing cases (90\%) have errors less than 0.05~ppm averaged over the simulation time steps. 
The remaining cases, as illustrated in the 90\% to 100\% distribution, have maximum errors larger than 0.3~ppm at the time steps after the first day. Despite the larger absolute error of these minority cases, the surrogate model remains numerically stable in all the testing cases. 
As the model is only trained on a small training dataset of two-day simulations and tested on a testing dataset of nine-day simulations, we conclude that the surrogate model captures the underlying system dynamics in most cases over a long period, but it might not generalize to conditions that were not included in the training data.

\subsection{Computational Speed}
We calculate the computational time using the Rosenbrock23 ODE solver for 3000 simulations over 48 hours of simulation time for both the reference chemical mechanism and the SINDy-based surrogate model. The total elapsed time for running 3000 simulations on the reference model is 83.7~s, and the time spent on the surrogate model with one, two, three, and four latent species are 2.9~s, 6.6~s, 10.4~s, and 27.0~s (reaching a speedup of 29$\times$, 13$\times$, 8$\times$, and 3$\times$, respectively). The surrogate models take fewer integration time steps. The reference model requires 976 steps on average for the ODE system integration, while the surrogate model with one, two, three, and four latent species requires 149, 339, 386, and 465 steps on average. 

\begin{figure}[h]
\centering
 \includegraphics[width=0.8\textwidth]{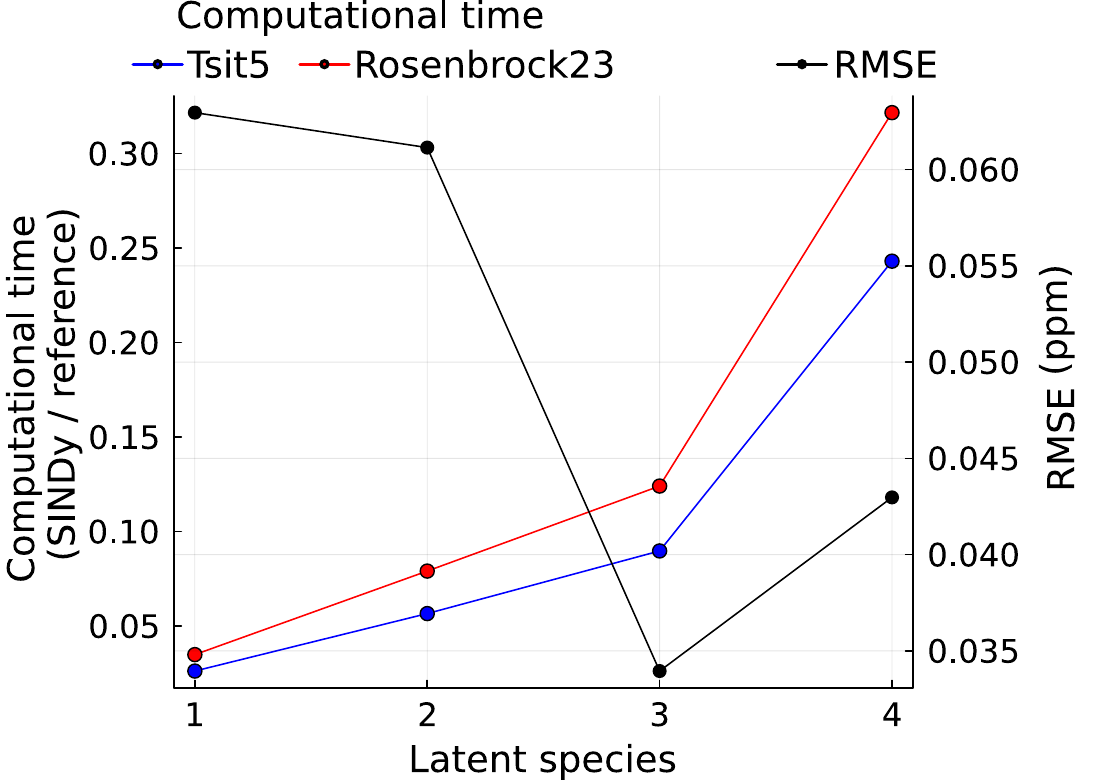}
 \caption{ML model computational speed and test set RMSE for ozone as functions of the number of
latent species. The red line compares the speed of the reference and ML models using the Rosenbrock23 solver; the blue line compares the reference model with the Rosenbrock23 solver to the ML models with the Tsit5 solver. (The reference model is too stiff for use with the Tsit5 solver.)}
 \label{fig:fig_CompSpeed}
 \end{figure}

As discussed in Section \ref{Stiffness Reduction}, our surrogate model is expected to be less stiff than the reference model, thereby allowing us to use solvers designed for non-stiff problems. Therefore, we also calculate the computational time using the Tsit5 ODE solver for 3000 simulations over 48 hours of simulation time for the surrogate model. For surrogate models with one, two, three, and four latent species, the total computational time are 2.2~s, 4.7~s, 7.5~s, and 20.3~s (reaching a speedup of 38$\times$, 18$\times$, 11$\times$, and 4$\times$, respectively), and the average numbers of integration timesteps are 31, 71, 82, and 106.
Computational speed ratios for the reference model vs. ML models are shown in Figure~\ref{fig:fig_CompSpeed}.

While surrogate models with smaller latent spaces are faster, there is a balance between the computational speed and the model performance (Figure~\ref{fig:fig_CompSpeed}).
Starting from one latent species, the RMSE decreases with more latent species, while the computational time increases approximately proportionally to  the number of terms on the RHS of the surrogate ODE(s), which increases nearly quadratically with the number of latent species (Figure~S11 to S14).  
Interestingly, adding the fourth latent species worsens the prediction of O\textsubscript{3}, probably due to the extra stiffness and redundancy that it adds, as the fourth latent species is not highly correlated to O\textsubscript{3} (Figure~\ref{fig:fig_LatentSpace}B). 
To meet the criteria of creating the fastest and most accurate model, we select the model with three latent species as a trade-off that is 11$\times$ faster and has 12$\times$ fewer integration timesteps than the reference model.

\section{Discussion and Conclusion}
In this work we have explored the feasibility of the SINDy approach to create surrogate models for atmospheric chemical mechanisms. 
Our SINDy-based model is numerically stable over the nine-day testing period under various environmental conditions. To our knowledge, previous works have either reported numerical instabilities or have not included experiments that quantify numerical stability, and we are the first to report numerical stability over long simulation periods under a wide range of conditions.
Our surrogate model achieves a 11$\times$ speedup compared to the reference mechanism in terms of the total integration time and requires 12$\times$ fewer integration timesteps  when using the fastest ODE solver for each system, and each timestep is 3.2$\times$ faster when both models use the same ODE solver. 
The surrogate model achieves such speedup because (1) the ODE system in the SINDy-based model has its dimensionality reduced to three state variables, and fewer state variables mean fewer computations; (2) we select a small and concise candidate function library based on prior understanding of the reference model so that the surrogate ODEs are highly simplified; and (3) because we use training data saved at one-hour intervals, the dynamics of the latent state variables all have similar characteristic time scales, allowing the surrogate model to be less numerically stiff than the reference model and to be solved by faster non-stiff ODE solvers such as Tsit5 (Section \ref{Computational Speed}). 

We reduce the dimensionality of the dataset from 11 chemical species into an interpretable latent space with three latent species. These latent species are linear combinations of chemical species that are either critical in the O\textsubscript{3} formation and destruction cycle or related to the pollutant emissions. The surrogate model representation is also a human-interpretable system of differential equations. As discussed above, the time derivatives of the latent species are highly dependent on the meteorological conditions (atmospheric pressure, temperature, and solar zenith angle) and emissions, indicating its potential ability for simulations under various environmental conditions. 

The results and findings herein represent certain step forward from previous literature; however, there is still more work to be done to achieve the full potential of this approach. 
First, the selection of terms for inclusion in the candidate function library is important as the performance of the SINDy-based surrogate model is sensitive to the choice of terms. 
This requires a deep prior understanding of the underlying dynamics in the reference chemical mechanism---a slight difference in the selected library terms might lead to substantially different surrogate dynamics which suffer from instabilities and failures. 
This difficulty may be exacerbated for larger-scale models, which contain more complex sub-mechanisms or multi-phase mechanisms such as inorganic and particulate matter physics and chemistry \cite{riemer2009simulating}, leading to more complicated dynamics than in the reference model used here.
However, the difficulty in choosing candidate terms may be mitigated by applying more robust variants or extensions of SINDy that are less sensitive to the selection of candidate functions such as E-SINDy \cite{fasel2022ensemble}, which uses bootstrapping techniques among the constructed library terms.
Second, some cases in the surrogate model simulations have occasional concentration spikes. We are able to prevent these from becoming exploding errors by using buffer terms as explained in Section \ref{Training Procedure}, but we are not yet able to prevent the spikes completely. This could be mitigated by using SINDy variants such as weak-SINDy \cite{messenger2021weak} which is trained on time-series integrations rather than individual time steps.

 In future work, we will scale up from the simplified mechanism used here to create compressed versions of the chemical mechanisms typically used in atmospheric chemical transport modeling, such as CB-6 \cite{yarwood2010cb6}, SAPRC \cite{carter1990saprc}, the GEOS-Chem mechanism \cite{bey2001geoschem} and even the Master Chemical Mechanism \cite{jenkin1997mcm1,saunders2003mcm2}. We will also expand our surrogates to cover aerosol dynamics and other atmospheric phenomena. By implementing the aforementioned methodology, we expect to obtain a larger speedup factor when creating surrogate models for a larger chemical mechanism as we hypothesize that larger chemical mechanisms include variables which are more strongly correlated with each other than exist in the small-scale mechanism used here.

%
%
%
%

%
%

%

%

\section{Open Research}
The source codes for dataset generation, model training, and plots are available through
Zenodo (\url{https://zenodo.org/records/10465785}).


\acknowledgments

This publication was developed with support from Grant No.~R840012
funded by the U.S. Environmental Protection Agency and an Early Career Faculty grant from the National Aeronautics and Space Administration (grant No.~80NSSC21K1813). It has not been formally reviewed by EPA or NASA. We are also grateful for computing resources provided by the National Center for Supercomputing Applications at the University of Illinois. The views expressed in this document are solely those of the authors. The authors declare no conflicts of interest relevant to this study.


%
%



\bibliography{agusample.bib}

\begin{thebibliography}{}

\bibitem [\protect \citeauthoryear {%
Ayesha%
, Hanif%
\BCBL {}\ \BBA {} Talib%
}{%
Ayesha%
\ \protect \BOthers {.}}{%
{\protect \APACyear {2020}}%
}]{%
ayesha2020overview}
\APACinsertmetastar {%
ayesha2020overview}%
\begin{APACrefauthors}%
Ayesha, S.%
, Hanif, M\BPBI K.%
\BCBL {}\ \BBA {} Talib, R.%
\end{APACrefauthors}%
\unskip\
\newblock
\APACrefYearMonthDay{2020}{}{}.
\newblock
{\BBOQ}\APACrefatitle {Overview and comparative study of dimensionality
  reduction techniques for high dimensional data} {Overview and comparative
  study of dimensionality reduction techniques for high dimensional
  data}.{\BBCQ}
\newblock
\APACjournalVolNumPages{{Information Fusion}}{59}{}{44-58}.
\newblock
\begin{APACrefDOI} \doi{https://doi.org/10.1016/j.inffus.2020.01.005}
  \end{APACrefDOI}
\PrintBackRefs{\CurrentBib}

\bibitem [\protect \citeauthoryear {%
Bagge~Carlson%
}{%
Bagge~Carlson%
}{%
{\protect \APACyear {2018}}%
}]{%
bagge2018hyperopt}
\APACinsertmetastar {%
bagge2018hyperopt}%
\begin{APACrefauthors}%
Bagge~Carlson, F.%
\end{APACrefauthors}%
\unskip\
\newblock
\APACrefYearMonthDay{2018}{August}{04}.
\newblock
\APACrefbtitle {Hyperopt.jl} {Hyperopt.jl}\ [Software].
\newblock
\APACaddressPublisher{}{Git{H}ub}.
\newblock
\begin{APACrefURL}
  \url{https://lup.lub.lu.se/search/publication/6ec19989-9b30-448c-be5e-bae4c4257c7b}
  \end{APACrefURL}
\PrintBackRefs{\CurrentBib}

\bibitem [\protect \citeauthoryear {%
Baxter%
}{%
Baxter%
}{%
{\protect \APACyear {2000}}%
}]{%
baxter2000model}
\APACinsertmetastar {%
baxter2000model}%
\begin{APACrefauthors}%
Baxter, J.%
\end{APACrefauthors}%
\unskip\
\newblock
\APACrefYearMonthDay{2000}{}{}.
\newblock
{\BBOQ}\APACrefatitle {A model of inductive bias learning} {A model of
  inductive bias learning}.{\BBCQ}
\newblock
\APACjournalVolNumPages{Journal of artificial intelligence
  research}{12}{}{149--198}.
\newblock
\begin{APACrefDOI} \doi{https://doi.org/10.1613/jair.731} \end{APACrefDOI}
\PrintBackRefs{\CurrentBib}

\bibitem [\protect \citeauthoryear {%
Bellman%
}{%
Bellman%
}{%
{\protect \APACyear {1966}}%
}]{%
bellman1966dynamic}
\APACinsertmetastar {%
bellman1966dynamic}%
\begin{APACrefauthors}%
Bellman, R.%
\end{APACrefauthors}%
\unskip\
\newblock
\APACrefYearMonthDay{1966}{}{}.
\newblock
{\BBOQ}\APACrefatitle {Dynamic Programming} {Dynamic programming}.{\BBCQ}
\newblock
\APACjournalVolNumPages{{Science}}{153}{3731}{34-37}.
\newblock
\begin{APACrefDOI} \doi{https://doi.org/10.1126/science.153.3731.34}
  \end{APACrefDOI}
\PrintBackRefs{\CurrentBib}

\bibitem [\protect \citeauthoryear {%
Bergstra%
\ \BBA {} Bengio%
}{%
Bergstra%
\ \BBA {} Bengio%
}{%
{\protect \APACyear {2012}}%
}]{%
bergstra2012random}
\APACinsertmetastar {%
bergstra2012random}%
\begin{APACrefauthors}%
Bergstra, J.%
\BCBT {}\ \BBA {} Bengio, Y.%
\end{APACrefauthors}%
\unskip\
\newblock
\APACrefYearMonthDay{2012}{}{}.
\newblock
{\BBOQ}\APACrefatitle {Random search for hyper-parameter optimization} {Random
  search for hyper-parameter optimization}.{\BBCQ}
\newblock
\APACjournalVolNumPages{{Journal of Machine Learning Research}}{13}{2}{}.
\newblock
\begin{APACrefDOI} \doi{https://dl.acm.org/doi/10.5555/2188385.2188395}
  \end{APACrefDOI}
\PrintBackRefs{\CurrentBib}

\bibitem [\protect \citeauthoryear {%
Bey%
\ \protect \BOthers {.}}{%
Bey%
\ \protect \BOthers {.}}{%
{\protect \APACyear {2001}}%
}]{%
bey2001geoschem}
\APACinsertmetastar {%
bey2001geoschem}%
\begin{APACrefauthors}%
Bey, I.%
, Jacob, D\BPBI J.%
, Yantosca, R\BPBI M.%
, Logan, J\BPBI A.%
, Field, B\BPBI D.%
, Fiore, A\BPBI M.%
\BDBL {}Schultz, M\BPBI G.%
\end{APACrefauthors}%
\unskip\
\newblock
\APACrefYearMonthDay{2001}{}{}.
\newblock
{\BBOQ}\APACrefatitle {Global modeling of tropospheric chemistry with
  assimilated meteorology: {M}odel description and evaluation} {Global modeling
  of tropospheric chemistry with assimilated meteorology: {M}odel description
  and evaluation}.{\BBCQ}
\newblock
\APACjournalVolNumPages{Journal of Geophysical Research:
  Atmospheres}{106}{D19}{23073--23095}.
\newblock
\begin{APACrefDOI} \doi{https://doi.org/10.1029/2001JD000807} \end{APACrefDOI}
\PrintBackRefs{\CurrentBib}

\bibitem [\protect \citeauthoryear {%
Bezanson%
, Edelman%
, Karpinski%
\BCBL {}\ \BBA {} Shah%
}{%
Bezanson%
\ \protect \BOthers {.}}{%
{\protect \APACyear {2017}}%
}]{%
julia}
\APACinsertmetastar {%
julia}%
\begin{APACrefauthors}%
Bezanson, J.%
, Edelman, A.%
, Karpinski, S.%
\BCBL {}\ \BBA {} Shah, V\BPBI B.%
\end{APACrefauthors}%
\unskip\
\newblock
\APACrefYearMonthDay{2017}{}{}.
\newblock
{\BBOQ}\APACrefatitle {Julia: {A} Fresh Approach to Numerical Computing}
  {Julia: {A} fresh approach to numerical computing}.{\BBCQ}
\newblock
\APACjournalVolNumPages{{SIAM Review}}{59}{1}{65-98}.
\newblock
\begin{APACrefDOI} \doi{https://doi.org/10.1137/141000671} \end{APACrefDOI}
\PrintBackRefs{\CurrentBib}

\bibitem [\protect \citeauthoryear {%
Boznar%
, Lesjak%
\BCBL {}\ \BBA {} Mlakar%
}{%
Boznar%
\ \protect \BOthers {.}}{%
{\protect \APACyear {1993}}%
}]{%
boznar1993nn}
\APACinsertmetastar {%
boznar1993nn}%
\begin{APACrefauthors}%
Boznar, M.%
, Lesjak, M.%
\BCBL {}\ \BBA {} Mlakar, P.%
\end{APACrefauthors}%
\unskip\
\newblock
\APACrefYearMonthDay{1993}{}{}.
\newblock
{\BBOQ}\APACrefatitle {A neural network-based method for short-term predictions
  of ambient {SO2} concentrations in highly polluted industrial areas of
  complex terrain} {A neural network-based method for short-term predictions of
  ambient {SO2} concentrations in highly polluted industrial areas of complex
  terrain}.{\BBCQ}
\newblock
\APACjournalVolNumPages{{Atmospheric Environment. Part B. Urban
  Atmosphere}}{27}{2}{221-230}.
\newblock
\begin{APACrefDOI} \doi{https://doi.org/10.1016/0957-1272(93)90007-S}
  \end{APACrefDOI}
\PrintBackRefs{\CurrentBib}

\bibitem [\protect \citeauthoryear {%
Brasseur%
\ \BBA {} Jacob%
}{%
Brasseur%
\ \BBA {} Jacob%
}{%
{\protect \APACyear {2017}}%
}]{%
brasseur2017AQMtextbook}
\APACinsertmetastar {%
brasseur2017AQMtextbook}%
\begin{APACrefauthors}%
Brasseur, G\BPBI P.%
\BCBT {}\ \BBA {} Jacob, D\BPBI J.%
\end{APACrefauthors}%
\unskip\
\newblock
\APACrefYear{2017}.
\newblock
\APACrefbtitle {Modeling of Atmospheric Chemistry} {Modeling of atmospheric
  chemistry}.
\newblock
\APACaddressPublisher{}{Cambridge University Press}.
\newblock
\begin{APACrefDOI} \doi{https://doi.org/10.1017/9781316544754} \end{APACrefDOI}
\PrintBackRefs{\CurrentBib}

\bibitem [\protect \citeauthoryear {%
Brunton%
, Noack%
\BCBL {}\ \BBA {} Koumoutsakos%
}{%
Brunton%
\ \protect \BOthers {.}}{%
{\protect \APACyear {2020}}%
}]{%
brunton2020fluid}
\APACinsertmetastar {%
brunton2020fluid}%
\begin{APACrefauthors}%
Brunton, S\BPBI L.%
, Noack, B\BPBI R.%
\BCBL {}\ \BBA {} Koumoutsakos, P.%
\end{APACrefauthors}%
\unskip\
\newblock
\APACrefYearMonthDay{2020}{}{}.
\newblock
{\BBOQ}\APACrefatitle {Machine Learning for Fluid Mechanics} {Machine learning
  for fluid mechanics}.{\BBCQ}
\newblock
\APACjournalVolNumPages{{Annual Review of Fluid Mechanics}}{52}{1}{477-508}.
\newblock
\begin{APACrefDOI} \doi{https://doi.org/10.1146/annurev-fluid-010719-060214}
  \end{APACrefDOI}
\PrintBackRefs{\CurrentBib}

\bibitem [\protect \citeauthoryear {%
Brunton%
, Proctor%
\BCBL {}\ \BBA {} Kutz%
}{%
Brunton%
\ \protect \BOthers {.}}{%
{\protect \APACyear {2016}}%
}]{%
brunton2016sindy}
\APACinsertmetastar {%
brunton2016sindy}%
\begin{APACrefauthors}%
Brunton, S\BPBI L.%
, Proctor, J\BPBI L.%
\BCBL {}\ \BBA {} Kutz, J\BPBI N.%
\end{APACrefauthors}%
\unskip\
\newblock
\APACrefYearMonthDay{2016}{}{}.
\newblock
{\BBOQ}\APACrefatitle {Discovering governing equations from data by sparse
  identification of nonlinear dynamical systems} {Discovering governing
  equations from data by sparse identification of nonlinear dynamical
  systems}.{\BBCQ}
\newblock
\APACjournalVolNumPages{{Proceedings of the National Academy of
  Sciences}}{113}{15}{3932-3937}.
\newblock
\begin{APACrefDOI} \doi{https://doi.org/10.1073/pnas.1517384113}
  \end{APACrefDOI}
\PrintBackRefs{\CurrentBib}

\bibitem [\protect \citeauthoryear {%
Cabaneros%
, Calautit%
\BCBL {}\ \BBA {} Hughes%
}{%
Cabaneros%
\ \protect \BOthers {.}}{%
{\protect \APACyear {2019}}%
}]{%
cabaneros2019review}
\APACinsertmetastar {%
cabaneros2019review}%
\begin{APACrefauthors}%
Cabaneros, S\BPBI M.%
, Calautit, J\BPBI K.%
\BCBL {}\ \BBA {} Hughes, B\BPBI R.%
\end{APACrefauthors}%
\unskip\
\newblock
\APACrefYearMonthDay{2019}{}{}.
\newblock
{\BBOQ}\APACrefatitle {A review of artificial neural network models for ambient
  air pollution prediction} {A review of artificial neural network models for
  ambient air pollution prediction}.{\BBCQ}
\newblock
\APACjournalVolNumPages{{Environmental Modelling \& Software}}{119}{}{285-304}.
\newblock
\begin{APACrefDOI} \doi{https://doi.org/10.1016/j.envsoft.2019.06.014}
  \end{APACrefDOI}
\PrintBackRefs{\CurrentBib}

\bibitem [\protect \citeauthoryear {%
Carter%
}{%
Carter%
}{%
{\protect \APACyear {1990}}%
}]{%
carter1990saprc}
\APACinsertmetastar {%
carter1990saprc}%
\begin{APACrefauthors}%
Carter, W\BPBI P.%
\end{APACrefauthors}%
\unskip\
\newblock
\APACrefYearMonthDay{1990}{}{}.
\newblock
{\BBOQ}\APACrefatitle {A detailed mechanism for the gas-phase atmospheric
  reactions of organic compounds} {A detailed mechanism for the gas-phase
  atmospheric reactions of organic compounds}.{\BBCQ}
\newblock
\APACjournalVolNumPages{{Atmospheric Environment. Part A. General
  Topics}}{24}{3}{481-518}.
\newblock
\begin{APACrefDOI} \doi{https://doi.org/10.1016/0960-1686(90)90005-8}
  \end{APACrefDOI}
\PrintBackRefs{\CurrentBib}

\bibitem [\protect \citeauthoryear {%
J.~Chen%
\ \BBA {} Revels%
}{%
J.~Chen%
\ \BBA {} Revels%
}{%
{\protect \APACyear {2016}}%
}]{%
BenchmarkTools2016}
\APACinsertmetastar {%
BenchmarkTools2016}%
\begin{APACrefauthors}%
Chen, J.%
\BCBT {}\ \BBA {} Revels, J.%
\end{APACrefauthors}%
\unskip\
\newblock
\APACrefYearMonthDay{2016}{}{}.
\newblock
{\BBOQ}\APACrefatitle {Robust benchmarking in noisy environments} {Robust
  benchmarking in noisy environments}.{\BBCQ}
\newblock
\APACjournalVolNumPages{arXiv}{}{}{}.
\newblock
\begin{APACrefDOI} \doi{https://doi.org/10.48550/arXiv.1608.04295}
  \end{APACrefDOI}
\PrintBackRefs{\CurrentBib}

\bibitem [\protect \citeauthoryear {%
R\BPBI T\BPBI Q.~Chen%
, Rubanova%
, Bettencourt%
\BCBL {}\ \BBA {} Duvenaud%
}{%
R\BPBI T\BPBI Q.~Chen%
\ \protect \BOthers {.}}{%
{\protect \APACyear {2019}}%
}]{%
chen2018node}
\APACinsertmetastar {%
chen2018node}%
\begin{APACrefauthors}%
Chen, R\BPBI T\BPBI Q.%
, Rubanova, Y.%
, Bettencourt, J.%
\BCBL {}\ \BBA {} Duvenaud, D.%
\end{APACrefauthors}%
\unskip\
\newblock
\APACrefYearMonthDay{2019}{}{}.
\newblock
{\BBOQ}\APACrefatitle {Neural Ordinary Differential Equations} {Neural ordinary
  differential equations}.{\BBCQ}
\newblock
\APACjournalVolNumPages{arXiv}{}{}{}.
\newblock
\begin{APACrefDOI} \doi{https://doi.org/10.48550/arXiv.1806.07366}
  \end{APACrefDOI}
\PrintBackRefs{\CurrentBib}

\bibitem [\protect \citeauthoryear {%
Djeumou%
, Neary%
, Goubault%
, Putot%
\BCBL {}\ \BBA {} Topcu%
}{%
Djeumou%
\ \protect \BOthers {.}}{%
{\protect \APACyear {2022}}%
}]{%
djeumou2022taylor}
\APACinsertmetastar {%
djeumou2022taylor}%
\begin{APACrefauthors}%
Djeumou, F.%
, Neary, C.%
, Goubault, {\'E}.%
, Putot, S.%
\BCBL {}\ \BBA {} Topcu, U.%
\end{APACrefauthors}%
\unskip\
\newblock
\APACrefYearMonthDay{2022}{}{}.
\newblock
{\BBOQ}\APACrefatitle {Taylor-Lagrange Neural Ordinary Differential Equations:
  Toward Fast Training and Evaluation of Neural ODEs} {Taylor-lagrange neural
  ordinary differential equations: Toward fast training and evaluation of
  neural odes}.{\BBCQ}
\newblock
\APACjournalVolNumPages{arXiv}{}{}{}.
\newblock
\begin{APACrefDOI} \doi{https://doi.org/10.24963/ijcai.2022/405}
  \end{APACrefDOI}
\PrintBackRefs{\CurrentBib}

\bibitem [\protect \citeauthoryear {%
Elman%
}{%
Elman%
}{%
{\protect \APACyear {1990}}%
}]{%
elman1990rnn}
\APACinsertmetastar {%
elman1990rnn}%
\begin{APACrefauthors}%
Elman, J\BPBI L.%
\end{APACrefauthors}%
\unskip\
\newblock
\APACrefYearMonthDay{1990}{}{}.
\newblock
{\BBOQ}\APACrefatitle {Finding structure in time} {Finding structure in
  time}.{\BBCQ}
\newblock
\APACjournalVolNumPages{{Cognitive Science}}{14}{2}{179-211}.
\newblock
\begin{APACrefDOI} \doi{https://doi.org/10.1016/0364-0213(90)90002-E}
  \end{APACrefDOI}
\PrintBackRefs{\CurrentBib}

\bibitem [\protect \citeauthoryear {%
Fasel%
, Kutz%
, Brunton%
\BCBL {}\ \BBA {} Brunton%
}{%
Fasel%
\ \protect \BOthers {.}}{%
{\protect \APACyear {2022}}%
}]{%
fasel2022ensemble}
\APACinsertmetastar {%
fasel2022ensemble}%
\begin{APACrefauthors}%
Fasel, U.%
, Kutz, J\BPBI N.%
, Brunton, B\BPBI W.%
\BCBL {}\ \BBA {} Brunton, S\BPBI L.%
\end{APACrefauthors}%
\unskip\
\newblock
\APACrefYearMonthDay{2022}{}{}.
\newblock
{\BBOQ}\APACrefatitle {Ensemble-SINDy: {R}obust sparse model discovery in the
  low-data, high-noise limit, with active learning and control}
  {Ensemble-sindy: {R}obust sparse model discovery in the low-data, high-noise
  limit, with active learning and control}.{\BBCQ}
\newblock
\APACjournalVolNumPages{{Proceedings of the Royal Society
  A}}{478}{2260}{20210904}.
\newblock
\begin{APACrefDOI} \doi{http://doi.org/10.1098/rspa.2021.0904} \end{APACrefDOI}
\PrintBackRefs{\CurrentBib}

\bibitem [\protect \citeauthoryear {%
Gardner%
\ \BBA {} Dorling%
}{%
Gardner%
\ \BBA {} Dorling%
}{%
{\protect \APACyear {1998}}%
}]{%
gardner1998nnreview}
\APACinsertmetastar {%
gardner1998nnreview}%
\begin{APACrefauthors}%
Gardner, M.%
\BCBT {}\ \BBA {} Dorling, S.%
\end{APACrefauthors}%
\unskip\
\newblock
\APACrefYearMonthDay{1998}{}{}.
\newblock
{\BBOQ}\APACrefatitle {Artificial neural networks (the multilayer
  perceptron)—{A} review of applications in the atmospheric sciences}
  {Artificial neural networks (the multilayer perceptron)—{A} review of
  applications in the atmospheric sciences}.{\BBCQ}
\newblock
\APACjournalVolNumPages{{Atmospheric Environment}}{32}{14}{2627-2636}.
\newblock
\begin{APACrefDOI} \doi{https://doi.org/10.1016/S1352-2310(97)00447-0}
  \end{APACrefDOI}
\PrintBackRefs{\CurrentBib}

\bibitem [\protect \citeauthoryear {%
Gery%
, Whitten%
, Killus%
\BCBL {}\ \BBA {} Dodge%
}{%
Gery%
\ \protect \BOthers {.}}{%
{\protect \APACyear {1989}}%
}]{%
gery1989cbm4}
\APACinsertmetastar {%
gery1989cbm4}%
\begin{APACrefauthors}%
Gery, M\BPBI W.%
, Whitten, G\BPBI Z.%
, Killus, J\BPBI P.%
\BCBL {}\ \BBA {} Dodge, M\BPBI C.%
\end{APACrefauthors}%
\unskip\
\newblock
\APACrefYearMonthDay{1989}{}{}.
\newblock
{\BBOQ}\APACrefatitle {A photochemical kinetics mechanism for urban and
  regional scale computer modeling} {A photochemical kinetics mechanism for
  urban and regional scale computer modeling}.{\BBCQ}
\newblock
\APACjournalVolNumPages{{Journal of Geophysical Research:
  Atmospheres}}{94}{D10}{12925-12956}.
\newblock
\begin{APACrefDOI} \doi{https://doi.org/10.1029/JD094iD10p12925}
  \end{APACrefDOI}
\PrintBackRefs{\CurrentBib}

\bibitem [\protect \citeauthoryear {%
Glynn%
}{%
Glynn%
}{%
{\protect \APACyear {2005}}%
}]{%
glynn2005initial}
\APACinsertmetastar {%
glynn2005initial}%
\begin{APACrefauthors}%
Glynn, P\BPBI W.%
\end{APACrefauthors}%
\unskip\
\newblock
\APACrefYearMonthDay{2005}{}{}.
\newblock
{\BBOQ}\APACrefatitle {Initial transient problem for steady-state output
  analysis} {Initial transient problem for steady-state output
  analysis}.{\BBCQ}
\newblock
\BIn{} \APACrefbtitle {Proceedings of the Winter Simulation Conference, 2005.}
  {Proceedings of the winter simulation conference, 2005.}
\newblock
\begin{APACrefDOI} \doi{https://doi.org/10.1109/WSC.2005.1574316}
  \end{APACrefDOI}
\PrintBackRefs{\CurrentBib}

\bibitem [\protect \citeauthoryear {%
Hirsh%
, Barajas-Solano%
\BCBL {}\ \BBA {} Kutz%
}{%
Hirsh%
\ \protect \BOthers {.}}{%
{\protect \APACyear {2022}}%
}]{%
hirsh2022sparsifying}
\APACinsertmetastar {%
hirsh2022sparsifying}%
\begin{APACrefauthors}%
Hirsh, S\BPBI M.%
, Barajas-Solano, D\BPBI A.%
\BCBL {}\ \BBA {} Kutz, J\BPBI N.%
\end{APACrefauthors}%
\unskip\
\newblock
\APACrefYearMonthDay{2022}{}{}.
\newblock
{\BBOQ}\APACrefatitle {Sparsifying priors for Bayesian uncertainty
  quantification in model discovery} {Sparsifying priors for bayesian
  uncertainty quantification in model discovery}.{\BBCQ}
\newblock
\APACjournalVolNumPages{{Royal Society Open Science}}{9}{2}{211823}.
\newblock
\begin{APACrefDOI} \doi{https://doi.org/10.1098/rsos.211823} \end{APACrefDOI}
\PrintBackRefs{\CurrentBib}

\bibitem [\protect \citeauthoryear {%
Hochreiter%
\ \BBA {} Schmidhuber%
}{%
Hochreiter%
\ \BBA {} Schmidhuber%
}{%
{\protect \APACyear {1997}}%
}]{%
hochreiter1997lstm}
\APACinsertmetastar {%
hochreiter1997lstm}%
\begin{APACrefauthors}%
Hochreiter, S.%
\BCBT {}\ \BBA {} Schmidhuber, J.%
\end{APACrefauthors}%
\unskip\
\newblock
\APACrefYearMonthDay{1997}{}{}.
\newblock
{\BBOQ}\APACrefatitle {Long short-term memory} {Long short-term memory}.{\BBCQ}
\newblock
\APACjournalVolNumPages{{Neural Computation}}{9}{8}{1735-1780}.
\newblock
\begin{APACrefDOI} \doi{https://doi.org/10.1162/neco.1997.9.8.1735}
  \end{APACrefDOI}
\PrintBackRefs{\CurrentBib}

\bibitem [\protect \citeauthoryear {%
Hoffmann%
, Fröhner%
\BCBL {}\ \BBA {} Noé%
}{%
Hoffmann%
\ \protect \BOthers {.}}{%
{\protect \APACyear {2019}}%
}]{%
hoffmann2019reactive}
\APACinsertmetastar {%
hoffmann2019reactive}%
\begin{APACrefauthors}%
Hoffmann, M.%
, Fröhner, C.%
\BCBL {}\ \BBA {} Noé, F.%
\end{APACrefauthors}%
\unskip\
\newblock
\APACrefYearMonthDay{2019}{}{}.
\newblock
{\BBOQ}\APACrefatitle {{Reactive SINDy: {D}iscovering governing reactions from
  concentration data}} {{Reactive SINDy: {D}iscovering governing reactions from
  concentration data}}.{\BBCQ}
\newblock
\APACjournalVolNumPages{The Journal of Chemical Physics}{150}{2}{025101}.
\newblock
\begin{APACrefDOI} \doi{https://doi.org/10.1063/1.5066099} \end{APACrefDOI}
\PrintBackRefs{\CurrentBib}

\bibitem [\protect \citeauthoryear {%
Hornik%
, Stinchcombe%
\BCBL {}\ \BBA {} White%
}{%
Hornik%
\ \protect \BOthers {.}}{%
{\protect \APACyear {1989}}%
}]{%
hornik1989nnapproximator}
\APACinsertmetastar {%
hornik1989nnapproximator}%
\begin{APACrefauthors}%
Hornik, K.%
, Stinchcombe, M.%
\BCBL {}\ \BBA {} White, H.%
\end{APACrefauthors}%
\unskip\
\newblock
\APACrefYearMonthDay{1989}{}{}.
\newblock
{\BBOQ}\APACrefatitle {Multilayer feedforward networks are universal
  approximators} {Multilayer feedforward networks are universal
  approximators}.{\BBCQ}
\newblock
\APACjournalVolNumPages{{Neural Networks}}{2}{5}{359-366}.
\newblock
\begin{APACrefDOI} \doi{https://doi.org/10.1016/0893-6080(89)90020-8}
  \end{APACrefDOI}
\PrintBackRefs{\CurrentBib}

\bibitem [\protect \citeauthoryear {%
Huang%
\ \BBA {} Seinfeld%
}{%
Huang%
\ \BBA {} Seinfeld%
}{%
{\protect \APACyear {2022}}%
}]{%
huang2022neural}
\APACinsertmetastar {%
huang2022neural}%
\begin{APACrefauthors}%
Huang, Y.%
\BCBT {}\ \BBA {} Seinfeld, J\BPBI H.%
\end{APACrefauthors}%
\unskip\
\newblock
\APACrefYearMonthDay{2022}{}{}.
\newblock
{\BBOQ}\APACrefatitle {A Neural Network-Assisted Euler Integrator for Stiff
  Kinetics in Atmospheric Chemistry} {A neural network-assisted euler
  integrator for stiff kinetics in atmospheric chemistry}.{\BBCQ}
\newblock
\APACjournalVolNumPages{{Environmental Science \&
  Technology}}{56}{7}{4676-4685}.
\newblock
\begin{APACrefDOI} \doi{https://doi.org/10.1021/acs.est.1c07648}
  \end{APACrefDOI}
\PrintBackRefs{\CurrentBib}

\bibitem [\protect \citeauthoryear {%
Jacobson%
}{%
Jacobson%
}{%
{\protect \APACyear {1997}}%
}]{%
jacobson1997effect}
\APACinsertmetastar {%
jacobson1997effect}%
\begin{APACrefauthors}%
Jacobson, S\BPBI H.%
\end{APACrefauthors}%
\unskip\
\newblock
\APACrefYearMonthDay{1997}{}{}.
\newblock
{\BBOQ}\APACrefatitle {The effect of initial transient on the steady-state
  simulation harmonic analysis gradient estimators} {The effect of initial
  transient on the steady-state simulation harmonic analysis gradient
  estimators}.{\BBCQ}
\newblock
\APACjournalVolNumPages{{Mathematics and Computers in
  Simulation}}{43}{2}{209-221}.
\newblock
\begin{APACrefDOI} \doi{https://doi.org/10.1016/S0378-4754(96)00068-7}
  \end{APACrefDOI}
\PrintBackRefs{\CurrentBib}

\bibitem [\protect \citeauthoryear {%
Jenkin%
, Saunders%
\BCBL {}\ \BBA {} Pilling%
}{%
Jenkin%
\ \protect \BOthers {.}}{%
{\protect \APACyear {1997}}%
}]{%
jenkin1997mcm1}
\APACinsertmetastar {%
jenkin1997mcm1}%
\begin{APACrefauthors}%
Jenkin, M\BPBI E.%
, Saunders, S\BPBI M.%
\BCBL {}\ \BBA {} Pilling, M\BPBI J.%
\end{APACrefauthors}%
\unskip\
\newblock
\APACrefYearMonthDay{1997}{}{}.
\newblock
{\BBOQ}\APACrefatitle {The tropospheric degradation of volatile organic
  compounds: a protocol for mechanism development} {The tropospheric
  degradation of volatile organic compounds: a protocol for mechanism
  development}.{\BBCQ}
\newblock
\APACjournalVolNumPages{{Atmospheric Environment}}{31}{1}{81-104}.
\newblock
\begin{APACrefDOI} \doi{https://doi.org/10.1016/S1352-2310(96)00105-7}
  \end{APACrefDOI}
\PrintBackRefs{\CurrentBib}

\bibitem [\protect \citeauthoryear {%
Jiang%
\ \protect \BOthers {.}}{%
Jiang%
\ \protect \BOthers {.}}{%
{\protect \APACyear {2021}}%
}]{%
jiang2021modeling}
\APACinsertmetastar {%
jiang2021modeling}%
\begin{APACrefauthors}%
Jiang, Y.%
, Xiong, X.%
, Zhang, S.%
, Wang, J.%
, Li, J.%
\BCBL {}\ \BBA {} Du, L.%
\end{APACrefauthors}%
\unskip\
\newblock
\APACrefYearMonthDay{2021}{}{}.
\newblock
{\BBOQ}\APACrefatitle {Modeling and prediction of the transmission dynamics of
  {COVID}-19 based on the {SIND}y-{LM} method} {Modeling and prediction of the
  transmission dynamics of {COVID}-19 based on the {SIND}y-{LM} method}.{\BBCQ}
\newblock
\APACjournalVolNumPages{Nonlinear Dynamics}{105}{3}{2775--2794}.
\newblock
\begin{APACrefDOI} \doi{https://doi.org/10.1007/s11071-021-06707-6}
  \end{APACrefDOI}
\PrintBackRefs{\CurrentBib}

\bibitem [\protect \citeauthoryear {%
Jolliffe%
\ \BBA {} Cadima%
}{%
Jolliffe%
\ \BBA {} Cadima%
}{%
{\protect \APACyear {2016}}%
}]{%
jolliffe2016principal}
\APACinsertmetastar {%
jolliffe2016principal}%
\begin{APACrefauthors}%
Jolliffe, I\BPBI T.%
\BCBT {}\ \BBA {} Cadima, J.%
\end{APACrefauthors}%
\unskip\
\newblock
\APACrefYearMonthDay{2016}{}{}.
\newblock
{\BBOQ}\APACrefatitle {Principal component analysis: {A} review and recent
  developments} {Principal component analysis: {A} review and recent
  developments}.{\BBCQ}
\newblock
\APACjournalVolNumPages{{Philosophical Transactions of the Royal Society A:
  Mathematical, Physical and Engineering Sciences}}{374}{2065}{20150202}.
\newblock
\begin{APACrefDOI} \doi{http://doi.org/10.1098/rsta.2015.0202} \end{APACrefDOI}
\PrintBackRefs{\CurrentBib}

\bibitem [\protect \citeauthoryear {%
Karniadakis%
\ \protect \BOthers {.}}{%
Karniadakis%
\ \protect \BOthers {.}}{%
{\protect \APACyear {2021}}%
}]{%
karniadakis2021piml}
\APACinsertmetastar {%
karniadakis2021piml}%
\begin{APACrefauthors}%
Karniadakis, G\BPBI E.%
, Kevrekidis, I\BPBI G.%
, Lu, L.%
, Perdikaris, P.%
, Wang, S.%
\BCBL {}\ \BBA {} Yang, L.%
\end{APACrefauthors}%
\unskip\
\newblock
\APACrefYearMonthDay{2021}{}{}.
\newblock
{\BBOQ}\APACrefatitle {Physics-informed machine learning} {Physics-informed
  machine learning}.{\BBCQ}
\newblock
\APACjournalVolNumPages{{Nature Reviews Physics}}{3}{6}{422--440}.
\newblock
\begin{APACrefDOI} \doi{https://doi.org/10.1038/s42254-021-00314-5}
  \end{APACrefDOI}
\PrintBackRefs{\CurrentBib}

\bibitem [\protect \citeauthoryear {%
Keller%
\ \BBA {} Evans%
}{%
Keller%
\ \BBA {} Evans%
}{%
{\protect \APACyear {2019}}%
}]{%
keller2019application}
\APACinsertmetastar {%
keller2019application}%
\begin{APACrefauthors}%
Keller, C\BPBI A.%
\BCBT {}\ \BBA {} Evans, M\BPBI J.%
\end{APACrefauthors}%
\unskip\
\newblock
\APACrefYearMonthDay{2019}{}{}.
\newblock
{\BBOQ}\APACrefatitle {Application of random forest regression to the
  calculation of gas-phase chemistry within the {GEOS}-{C}hem chemistry model
  v10} {Application of random forest regression to the calculation of gas-phase
  chemistry within the {GEOS}-{C}hem chemistry model v10}.{\BBCQ}
\newblock
\APACjournalVolNumPages{{Geoscientific Model Development}}{12}{3}{1209--1225}.
\newblock
\begin{APACrefDOI} \doi{https://doi.org/10.5194/gmd-12-1209-2019}
  \end{APACrefDOI}
\PrintBackRefs{\CurrentBib}

\bibitem [\protect \citeauthoryear {%
Kelp%
, Jacob%
, Kutz%
, Marshall%
\BCBL {}\ \BBA {} Tessum%
}{%
Kelp%
\ \protect \BOthers {.}}{%
{\protect \APACyear {2020}}%
}]{%
kelp2020rnn}
\APACinsertmetastar {%
kelp2020rnn}%
\begin{APACrefauthors}%
Kelp, M\BPBI M.%
, Jacob, D\BPBI J.%
, Kutz, J\BPBI N.%
, Marshall, J\BPBI D.%
\BCBL {}\ \BBA {} Tessum, C\BPBI W.%
\end{APACrefauthors}%
\unskip\
\newblock
\APACrefYearMonthDay{2020}{}{}.
\newblock
{\BBOQ}\APACrefatitle {Toward Stable, General Machine-Learned Models of the
  Atmospheric Chemical System} {Toward stable, general machine-learned models
  of the atmospheric chemical system}.{\BBCQ}
\newblock
\APACjournalVolNumPages{{Journal of Geophysical Research:
  Atmospheres}}{125}{23}{e2020JD032759}.
\newblock
\begin{APACrefDOI} \doi{https://doi.org/10.1029/2020JD032759} \end{APACrefDOI}
\PrintBackRefs{\CurrentBib}

\bibitem [\protect \citeauthoryear {%
Kelp%
, Tessum%
\BCBL {}\ \BBA {} Marshall%
}{%
Kelp%
\ \protect \BOthers {.}}{%
{\protect \APACyear {2018}}%
}]{%
kelp2018orders}
\APACinsertmetastar {%
kelp2018orders}%
\begin{APACrefauthors}%
Kelp, M\BPBI M.%
, Tessum, C\BPBI W.%
\BCBL {}\ \BBA {} Marshall, J\BPBI D.%
\end{APACrefauthors}%
\unskip\
\newblock
\APACrefYearMonthDay{2018}{}{}.
\newblock
{\BBOQ}\APACrefatitle {Orders-of-magnitude speedup in atmospheric chemistry
  modeling through neural network-based emulation} {Orders-of-magnitude speedup
  in atmospheric chemistry modeling through neural network-based
  emulation}.{\BBCQ}
\newblock
\APACjournalVolNumPages{arXiv}{}{}{}.
\newblock
\begin{APACrefDOI} \doi{https://doi.org/10.48550/arXiv.1808.03874}
  \end{APACrefDOI}
\PrintBackRefs{\CurrentBib}

\bibitem [\protect \citeauthoryear {%
Kim%
, Ji%
, Deng%
, Ma%
\BCBL {}\ \BBA {} Rackauckas%
}{%
Kim%
\ \protect \BOthers {.}}{%
{\protect \APACyear {2021}}%
}]{%
kim2021stiff}
\APACinsertmetastar {%
kim2021stiff}%
\begin{APACrefauthors}%
Kim, S.%
, Ji, W.%
, Deng, S.%
, Ma, Y.%
\BCBL {}\ \BBA {} Rackauckas, C.%
\end{APACrefauthors}%
\unskip\
\newblock
\APACrefYearMonthDay{2021}{}{}.
\newblock
{\BBOQ}\APACrefatitle {Stiff neural ordinary differential equations} {Stiff
  neural ordinary differential equations}.{\BBCQ}
\newblock
\APACjournalVolNumPages{Chaos: An Interdisciplinary Journal of Nonlinear
  Science}{31}{9}{093122}.
\newblock
\begin{APACrefDOI} \doi{https://doi.org/10.1063/5.0060697} \end{APACrefDOI}
\PrintBackRefs{\CurrentBib}

\bibitem [\protect \citeauthoryear {%
Kolen%
\ \BBA {} Kremer%
}{%
Kolen%
\ \BBA {} Kremer%
}{%
{\protect \APACyear {2001}}%
}]{%
hochreiter2001nn}
\APACinsertmetastar {%
hochreiter2001nn}%
\begin{APACrefauthors}%
Kolen, J\BPBI F.%
\BCBT {}\ \BBA {} Kremer, S\BPBI C.%
\end{APACrefauthors}%
\unskip\
\newblock
\APACrefYearMonthDay{2001}{}{}.
\newblock
{\BBOQ}\APACrefatitle {Gradient Flow in Recurrent Nets: The Difficulty of
  Learning LongTerm Dependencies} {Gradient flow in recurrent nets: The
  difficulty of learning longterm dependencies}.{\BBCQ}
\newblock
\BIn{} \APACrefbtitle {A Field Guide to Dynamical Recurrent Networks} {A field
  guide to dynamical recurrent networks}\ (\BPG~237-243).
\newblock
\APACaddressPublisher{}{IEEE}.
\newblock
\begin{APACrefDOI} \doi{https://doi.org/10.1109/9780470544037.ch14}
  \end{APACrefDOI}
\PrintBackRefs{\CurrentBib}

\bibitem [\protect \citeauthoryear {%
Kramer%
}{%
Kramer%
}{%
{\protect \APACyear {1991}}%
}]{%
kramer1991autoencoder}
\APACinsertmetastar {%
kramer1991autoencoder}%
\begin{APACrefauthors}%
Kramer, M\BPBI A.%
\end{APACrefauthors}%
\unskip\
\newblock
\APACrefYearMonthDay{1991}{}{}.
\newblock
{\BBOQ}\APACrefatitle {Nonlinear principal component analysis using
  autoassociative neural networks} {Nonlinear principal component analysis
  using autoassociative neural networks}.{\BBCQ}
\newblock
\APACjournalVolNumPages{{AIChE Journal}}{37}{2}{233-243}.
\newblock
\begin{APACrefDOI} \doi{https://doi.org/10.1002/aic.690370209} \end{APACrefDOI}
\PrintBackRefs{\CurrentBib}

\bibitem [\protect \citeauthoryear {%
Lai%
\ \BBA {} Nagarajaiah%
}{%
Lai%
\ \BBA {} Nagarajaiah%
}{%
{\protect \APACyear {2019}}%
}]{%
lai2019sparse}
\APACinsertmetastar {%
lai2019sparse}%
\begin{APACrefauthors}%
Lai, Z.%
\BCBT {}\ \BBA {} Nagarajaiah, S.%
\end{APACrefauthors}%
\unskip\
\newblock
\APACrefYearMonthDay{2019}{}{}.
\newblock
{\BBOQ}\APACrefatitle {Sparse structural system identification method for
  nonlinear dynamic systems with hysteresis/inelastic behavior} {Sparse
  structural system identification method for nonlinear dynamic systems with
  hysteresis/inelastic behavior}.{\BBCQ}
\newblock
\APACjournalVolNumPages{Mechanical Systems and Signal
  Processing}{117}{}{813-842}.
\newblock
\begin{APACrefDOI} \doi{https://doi.org/10.1016/j.ymssp.2018.08.033}
  \end{APACrefDOI}
\PrintBackRefs{\CurrentBib}

\bibitem [\protect \citeauthoryear {%
Logan%
, Prather%
, Wofsy%
\BCBL {}\ \BBA {} McElroy%
}{%
Logan%
\ \protect \BOthers {.}}{%
{\protect \APACyear {1981}}%
}]{%
logan1981multiscale}
\APACinsertmetastar {%
logan1981multiscale}%
\begin{APACrefauthors}%
Logan, J\BPBI A.%
, Prather, M\BPBI J.%
, Wofsy, S\BPBI C.%
\BCBL {}\ \BBA {} McElroy, M\BPBI B.%
\end{APACrefauthors}%
\unskip\
\newblock
\APACrefYearMonthDay{1981}{}{}.
\newblock
{\BBOQ}\APACrefatitle {Tropospheric chemistry: {A} global perspective}
  {Tropospheric chemistry: {A} global perspective}.{\BBCQ}
\newblock
\APACjournalVolNumPages{Journal of Geophysical Research:
  Oceans}{86}{C8}{7210-7254}.
\newblock
\begin{APACrefDOI} \doi{https://doi.org/10.1029/JC086iC08p07210}
  \end{APACrefDOI}
\PrintBackRefs{\CurrentBib}

\bibitem [\protect \citeauthoryear {%
Loman%
\ \protect \BOthers {.}}{%
Loman%
\ \protect \BOthers {.}}{%
{\protect \APACyear {2022}}%
}]{%
2022Catalyst}
\APACinsertmetastar {%
2022Catalyst}%
\begin{APACrefauthors}%
Loman, T\BPBI E.%
, Ma, Y.%
, Ilin, V.%
, Gowda, S.%
, Korsbo, N.%
, Yewale, N.%
\BDBL {}Isaacson, S\BPBI A.%
\end{APACrefauthors}%
\unskip\
\newblock
\APACrefYearMonthDay{2022}{}{}.
\newblock
{\BBOQ}\APACrefatitle {Catalyst: {F}ast Biochemical Modeling with {J}ulia}
  {Catalyst: {F}ast biochemical modeling with {J}ulia}.{\BBCQ}
\newblock
\APACjournalVolNumPages{bioRxiv}{}{}{}.
\newblock
\begin{APACrefDOI} \doi{https://doi.org/10.1101/2022.07.30.502135}
  \end{APACrefDOI}
\PrintBackRefs{\CurrentBib}

\bibitem [\protect \citeauthoryear {%
Ma%
\ \protect \BOthers {.}}{%
Ma%
\ \protect \BOthers {.}}{%
{\protect \APACyear {2022}}%
}]{%
ma2021modelingtoolkit}
\APACinsertmetastar {%
ma2021modelingtoolkit}%
\begin{APACrefauthors}%
Ma, Y.%
, Gowda, S.%
, Anantharaman, R.%
, Laughman, C.%
, Shah, V.%
\BCBL {}\ \BBA {} Rackauckas, C.%
\end{APACrefauthors}%
\unskip\
\newblock
\APACrefYearMonthDay{2022}{}{}.
\newblock
{\BBOQ}\APACrefatitle {Modeling{T}oolkit: {A} Composable Graph Transformation
  System For Equation-Based Modeling} {Modeling{T}oolkit: {A} composable graph
  transformation system for equation-based modeling}.{\BBCQ}
\newblock
\APACjournalVolNumPages{arXiv}{}{}{}.
\newblock
\begin{APACrefDOI} \doi{https://doi.org/10.48550/arXiv.2103.05244}
  \end{APACrefDOI}
\PrintBackRefs{\CurrentBib}

\bibitem [\protect \citeauthoryear {%
Martensen%
\ \protect \BOthers {.}}{%
Martensen%
\ \protect \BOthers {.}}{%
{\protect \APACyear {2023}}%
}]{%
datadrivendiffeq}
\APACinsertmetastar {%
datadrivendiffeq}%
\begin{APACrefauthors}%
Martensen, J.%
, Rackauckas, C.%
, Abrevaya, G.%
, Lee, G.%
, Strouwen, A.%
, Gwóźdź, M.%
\BDBL {}Štěpán Zapadlo%
\end{APACrefauthors}%
\unskip\
\newblock
\APACrefYearMonthDay{2023}{}{}.
\newblock
\APACrefbtitle {SciML/{D}ata{D}riven{D}iff{E}q.jl: v1.3.0}
  {Sciml/{D}ata{D}riven{D}iff{E}q.jl: v1.3.0}\ [Software].
\newblock
\APACaddressPublisher{}{Zenodo}.
\newblock
\begin{APACrefDOI} \doi{https://doi.org/10.5281/zenodo.8356369}
  \end{APACrefDOI}
\PrintBackRefs{\CurrentBib}

\bibitem [\protect \citeauthoryear {%
Maćkiewicz%
\ \BBA {} Ratajczak%
}{%
Maćkiewicz%
\ \BBA {} Ratajczak%
}{%
{\protect \APACyear {1993}}%
}]{%
MACKIEWICZ1993303}
\APACinsertmetastar {%
MACKIEWICZ1993303}%
\begin{APACrefauthors}%
Maćkiewicz, A.%
\BCBT {}\ \BBA {} Ratajczak, W.%
\end{APACrefauthors}%
\unskip\
\newblock
\APACrefYearMonthDay{1993}{}{}.
\newblock
{\BBOQ}\APACrefatitle {Principal components analysis ({PCA})} {Principal
  components analysis ({PCA})}.{\BBCQ}
\newblock
\APACjournalVolNumPages{Computers \& Geosciences}{19}{3}{303-342}.
\newblock
\begin{APACrefDOI} \doi{https://doi.org/10.1016/0098-3004(93)90090-R}
  \end{APACrefDOI}
\PrintBackRefs{\CurrentBib}

\bibitem [\protect \citeauthoryear {%
Messenger%
\ \BBA {} Bortz%
}{%
Messenger%
\ \BBA {} Bortz%
}{%
{\protect \APACyear {2021}}%
}]{%
messenger2021weak}
\APACinsertmetastar {%
messenger2021weak}%
\begin{APACrefauthors}%
Messenger, D\BPBI A.%
\BCBT {}\ \BBA {} Bortz, D\BPBI M.%
\end{APACrefauthors}%
\unskip\
\newblock
\APACrefYearMonthDay{2021}{}{}.
\newblock
{\BBOQ}\APACrefatitle {Weak SINDy: {G}alerkin-Based Data-Driven Model
  Selection} {Weak sindy: {G}alerkin-based data-driven model selection}.{\BBCQ}
\newblock
\APACjournalVolNumPages{{Multiscale Modeling \& Simulation}}{19}{3}{1474-1497}.
\newblock
\begin{APACrefDOI} \doi{https://doi.org/10.1137/20M134316} \end{APACrefDOI}
\PrintBackRefs{\CurrentBib}

\bibitem [\protect \citeauthoryear {%
Pasquato%
\ \protect \BOthers {.}}{%
Pasquato%
\ \protect \BOthers {.}}{%
{\protect \APACyear {2022}}%
}]{%
pasquato2022sparse}
\APACinsertmetastar {%
pasquato2022sparse}%
\begin{APACrefauthors}%
Pasquato, M.%
, Abbas, M.%
, Trani, A\BPBI A.%
, Nori, M.%
, Kwiecinski, J\BPBI A.%
, Trevisan, P.%
\BDBL {}Macciò, A\BPBI V.%
\end{APACrefauthors}%
\unskip\
\newblock
\APACrefYearMonthDay{2022}{}{}.
\newblock
{\BBOQ}\APACrefatitle {Sparse Identification of Variable Star Dynamics} {Sparse
  identification of variable star dynamics}.{\BBCQ}
\newblock
\APACjournalVolNumPages{The Astrophysical Journal}{930}{2}{161}.
\newblock
\begin{APACrefDOI} \doi{https://doi.org/10.3847/1538-4357/ac5624}
  \end{APACrefDOI}
\PrintBackRefs{\CurrentBib}

\bibitem [\protect \citeauthoryear {%
Purnomo%
\ \BBA {} Hayashibe%
}{%
Purnomo%
\ \BBA {} Hayashibe%
}{%
{\protect \APACyear {2023}}%
}]{%
purnomo2023sparse}
\APACinsertmetastar {%
purnomo2023sparse}%
\begin{APACrefauthors}%
Purnomo, A.%
\BCBT {}\ \BBA {} Hayashibe, M.%
\end{APACrefauthors}%
\unskip\
\newblock
\APACrefYearMonthDay{2023}{}{}.
\newblock
{\BBOQ}\APACrefatitle {Sparse identification of Lagrangian for nonlinear
  dynamical systems via proximal gradient method} {Sparse identification of
  lagrangian for nonlinear dynamical systems via proximal gradient
  method}.{\BBCQ}
\newblock
\APACjournalVolNumPages{{Scientific Reports}}{13}{1}{7919}.
\newblock
\begin{APACrefDOI} \doi{https://doi.org/10.1038/s41598-023-34931-0}
  \end{APACrefDOI}
\PrintBackRefs{\CurrentBib}

\bibitem [\protect \citeauthoryear {%
Quade%
, Abel%
, Shafi%
, Niven%
\BCBL {}\ \BBA {} Noack%
}{%
Quade%
\ \protect \BOthers {.}}{%
{\protect \APACyear {2016}}%
}]{%
quade2016prediction}
\APACinsertmetastar {%
quade2016prediction}%
\begin{APACrefauthors}%
Quade, M.%
, Abel, M.%
, Shafi, K.%
, Niven, R\BPBI K.%
\BCBL {}\ \BBA {} Noack, B\BPBI R.%
\end{APACrefauthors}%
\unskip\
\newblock
\APACrefYearMonthDay{2016}{}{}.
\newblock
{\BBOQ}\APACrefatitle {Prediction of dynamical systems by symbolic regression}
  {Prediction of dynamical systems by symbolic regression}.{\BBCQ}
\newblock
\APACjournalVolNumPages{{Physical Review E}}{94}{}{012214}.
\newblock
\begin{APACrefDOI} \doi{https://doi.org/10.1103/PhysRevE.94.012214}
  \end{APACrefDOI}
\PrintBackRefs{\CurrentBib}

\bibitem [\protect \citeauthoryear {%
Rackauckas%
\ \BBA {} Nie%
}{%
Rackauckas%
\ \BBA {} Nie%
}{%
{\protect \APACyear {2017}}%
}]{%
rackauckas2017differentialequations}
\APACinsertmetastar {%
rackauckas2017differentialequations}%
\begin{APACrefauthors}%
Rackauckas, C.%
\BCBT {}\ \BBA {} Nie, Q.%
\end{APACrefauthors}%
\unskip\
\newblock
\APACrefYearMonthDay{2017}{}{}.
\newblock
{\BBOQ}\APACrefatitle {Differential{E}quations.jl--{A} performant and
  feature-rich ecosystem for solving differential equations in {J}ulia}
  {Differential{E}quations.jl--{A} performant and feature-rich ecosystem for
  solving differential equations in {J}ulia}.{\BBCQ}
\newblock
\APACjournalVolNumPages{{Journal of Open Research Software}}{5}{1}{}.
\newblock
\begin{APACrefDOI} \doi{https://doi.org/10.5334/jors.151} \end{APACrefDOI}
\PrintBackRefs{\CurrentBib}

\bibitem [\protect \citeauthoryear {%
Rackauckas%
\ \BBA {} Nie%
}{%
Rackauckas%
\ \BBA {} Nie%
}{%
{\protect \APACyear {2019}}%
}]{%
rackauckas2019confederated}
\APACinsertmetastar {%
rackauckas2019confederated}%
\begin{APACrefauthors}%
Rackauckas, C.%
\BCBT {}\ \BBA {} Nie, Q.%
\end{APACrefauthors}%
\unskip\
\newblock
\APACrefYearMonthDay{2019}{}{}.
\newblock
{\BBOQ}\APACrefatitle {Confederated modular differential equation {API}s for
  accelerated algorithm development and benchmarking} {Confederated modular
  differential equation {API}s for accelerated algorithm development and
  benchmarking}.{\BBCQ}
\newblock
\APACjournalVolNumPages{Advances in Engineering Software}{132}{}{1-6}.
\newblock
\begin{APACrefDOI} \doi{https://doi.org/10.1016/j.advengsoft.2019.03.009}
  \end{APACrefDOI}
\PrintBackRefs{\CurrentBib}

\bibitem [\protect \citeauthoryear {%
Reichstein%
\ \protect \BOthers {.}}{%
Reichstein%
\ \protect \BOthers {.}}{%
{\protect \APACyear {2019}}%
}]{%
reichstein2019pinn}
\APACinsertmetastar {%
reichstein2019pinn}%
\begin{APACrefauthors}%
Reichstein, M.%
, Camps-Valls, G.%
, Stevens, B.%
, Jung, M.%
, Denzler, J.%
, Carvalhais, N.%
\BCBL {}\ \BBA {} Prabhat.%
\end{APACrefauthors}%
\unskip\
\newblock
\APACrefYearMonthDay{2019}{}{}.
\newblock
{\BBOQ}\APACrefatitle {Deep learning and process understanding for data-driven
  {E}arth system science} {Deep learning and process understanding for
  data-driven {E}arth system science}.{\BBCQ}
\newblock
\APACjournalVolNumPages{{Nature}}{566}{7743}{195--204}.
\newblock
\begin{APACrefDOI} \doi{https://doi.org/10.1038/s41586-019-0912-1}
  \end{APACrefDOI}
\PrintBackRefs{\CurrentBib}

\bibitem [\protect \citeauthoryear {%
Riemer%
, West%
, Zaveri%
\BCBL {}\ \BBA {} Easter%
}{%
Riemer%
\ \protect \BOthers {.}}{%
{\protect \APACyear {2009}}%
}]{%
riemer2009simulating}
\APACinsertmetastar {%
riemer2009simulating}%
\begin{APACrefauthors}%
Riemer, N.%
, West, M.%
, Zaveri, R\BPBI A.%
\BCBL {}\ \BBA {} Easter, R\BPBI C.%
\end{APACrefauthors}%
\unskip\
\newblock
\APACrefYearMonthDay{2009}{}{}.
\newblock
{\BBOQ}\APACrefatitle {Simulating the evolution of soot mixing state with a
  particle-resolved aerosol model} {Simulating the evolution of soot mixing
  state with a particle-resolved aerosol model}.{\BBCQ}
\newblock
\APACjournalVolNumPages{{Journal of Geophysical Research:
  Atmospheres}}{114}{D9}{}.
\newblock
\begin{APACrefDOI} \doi{https://doi.org/10.1029/2008JD011073} \end{APACrefDOI}
\PrintBackRefs{\CurrentBib}

\bibitem [\protect \citeauthoryear {%
Saunders%
, Jenkin%
, Derwent%
\BCBL {}\ \BBA {} Pilling%
}{%
Saunders%
\ \protect \BOthers {.}}{%
{\protect \APACyear {2003}}%
}]{%
saunders2003mcm2}
\APACinsertmetastar {%
saunders2003mcm2}%
\begin{APACrefauthors}%
Saunders, S\BPBI M.%
, Jenkin, M\BPBI E.%
, Derwent, R\BPBI G.%
\BCBL {}\ \BBA {} Pilling, M\BPBI J.%
\end{APACrefauthors}%
\unskip\
\newblock
\APACrefYearMonthDay{2003}{}{}.
\newblock
{\BBOQ}\APACrefatitle {Protocol for the development of the {M}aster {C}hemical
  {M}echanism, {MCM} v3 ({P}art {A}): {T}ropospheric degradation of
  non-aromatic volatile organic compounds} {Protocol for the development of the
  {M}aster {C}hemical {M}echanism, {MCM} v3 ({P}art {A}): {T}ropospheric
  degradation of non-aromatic volatile organic compounds}.{\BBCQ}
\newblock
\APACjournalVolNumPages{{Atmospheric Chemistry and Physics}}{3}{1}{161--180}.
\newblock
\begin{APACrefDOI} \doi{https://doi.org/10.5194/acp-3-161-2003}
  \end{APACrefDOI}
\PrintBackRefs{\CurrentBib}

\bibitem [\protect \citeauthoryear {%
Schmidt%
\ \BBA {} Lipson%
}{%
Schmidt%
\ \BBA {} Lipson%
}{%
{\protect \APACyear {2009}}%
}]{%
schmidt2009distilling}
\APACinsertmetastar {%
schmidt2009distilling}%
\begin{APACrefauthors}%
Schmidt, M.%
\BCBT {}\ \BBA {} Lipson, H.%
\end{APACrefauthors}%
\unskip\
\newblock
\APACrefYearMonthDay{2009}{}{}.
\newblock
{\BBOQ}\APACrefatitle {Distilling Free-Form Natural Laws from Experimental
  Data} {Distilling free-form natural laws from experimental data}.{\BBCQ}
\newblock
\APACjournalVolNumPages{{Science}}{324}{5923}{81-85}.
\newblock
\begin{APACrefDOI} \doi{https://doi.org/10.1126/science.1165893}
  \end{APACrefDOI}
\PrintBackRefs{\CurrentBib}

\bibitem [\protect \citeauthoryear {%
Shampine%
\ \BBA {} Reichelt%
}{%
Shampine%
\ \BBA {} Reichelt%
}{%
{\protect \APACyear {1997}}%
}]{%
shampine1997matlab}
\APACinsertmetastar {%
shampine1997matlab}%
\begin{APACrefauthors}%
Shampine, L\BPBI F.%
\BCBT {}\ \BBA {} Reichelt, M\BPBI W.%
\end{APACrefauthors}%
\unskip\
\newblock
\APACrefYearMonthDay{1997}{}{}.
\newblock
{\BBOQ}\APACrefatitle {The {MATLAB} {ODE} Suite} {The {MATLAB} {ODE}
  suite}.{\BBCQ}
\newblock
\APACjournalVolNumPages{SIAM Journal on Scientific Computing}{18}{1}{1-22}.
\newblock
\begin{APACrefDOI} \doi{https://doi.org/10.1137/S1064827594276424}
  \end{APACrefDOI}
\PrintBackRefs{\CurrentBib}

\bibitem [\protect \citeauthoryear {%
Sorjamaa%
, Hao%
, Reyhani%
, Ji%
\BCBL {}\ \BBA {} Lendasse%
}{%
Sorjamaa%
\ \protect \BOthers {.}}{%
{\protect \APACyear {2007}}%
}]{%
sorjamaa2007longtermtimeseriespred}
\APACinsertmetastar {%
sorjamaa2007longtermtimeseriespred}%
\begin{APACrefauthors}%
Sorjamaa, A.%
, Hao, J.%
, Reyhani, N.%
, Ji, Y.%
\BCBL {}\ \BBA {} Lendasse, A.%
\end{APACrefauthors}%
\unskip\
\newblock
\APACrefYearMonthDay{2007}{}{}.
\newblock
{\BBOQ}\APACrefatitle {Methodology for long-term prediction of time series}
  {Methodology for long-term prediction of time series}.{\BBCQ}
\newblock
\APACjournalVolNumPages{{Neurocomputing}}{70}{16}{2861-2869}.
\newblock
\begin{APACrefDOI} \doi{https://doi.org/10.1016/j.neucom.2006.06.015}
  \end{APACrefDOI}
\PrintBackRefs{\CurrentBib}

\bibitem [\protect \citeauthoryear {%
Sousa%
, Martins%
, Alvim-Ferraz%
\BCBL {}\ \BBA {} Pereira%
}{%
Sousa%
\ \protect \BOthers {.}}{%
{\protect \APACyear {2007}}%
}]{%
sousa2007pcann}
\APACinsertmetastar {%
sousa2007pcann}%
\begin{APACrefauthors}%
Sousa, S.%
, Martins, F.%
, Alvim-Ferraz, M.%
\BCBL {}\ \BBA {} Pereira, M.%
\end{APACrefauthors}%
\unskip\
\newblock
\APACrefYearMonthDay{2007}{}{}.
\newblock
{\BBOQ}\APACrefatitle {Multiple linear regression and artificial neural
  networks based on principal components to predict ozone concentrations}
  {Multiple linear regression and artificial neural networks based on principal
  components to predict ozone concentrations}.{\BBCQ}
\newblock
\APACjournalVolNumPages{{Environmental Modelling \& Software}}{22}{1}{97-103}.
\newblock
\begin{APACrefDOI} \doi{https://doi.org/10.1016/j.envsoft.2005.12.002}
  \end{APACrefDOI}
\PrintBackRefs{\CurrentBib}

\bibitem [\protect \citeauthoryear {%
Srivastava%
, Hinton%
, Krizhevsky%
, Sutskever%
\BCBL {}\ \BBA {} Salakhutdinov%
}{%
Srivastava%
\ \protect \BOthers {.}}{%
{\protect \APACyear {2014}}%
}]{%
srivastava2014dropout}
\APACinsertmetastar {%
srivastava2014dropout}%
\begin{APACrefauthors}%
Srivastava, N.%
, Hinton, G.%
, Krizhevsky, A.%
, Sutskever, I.%
\BCBL {}\ \BBA {} Salakhutdinov, R.%
\end{APACrefauthors}%
\unskip\
\newblock
\APACrefYearMonthDay{2014}{}{}.
\newblock
{\BBOQ}\APACrefatitle {Dropout: {A} Simple Way to Prevent Neural Networks from
  Overfitting} {Dropout: {A} simple way to prevent neural networks from
  overfitting}.{\BBCQ}
\newblock
\APACjournalVolNumPages{Journal of Machine Learning
  Research}{15}{56}{1929--1958}.
\newblock
\begin{APACrefDOI} \doi{https://dl.acm.org/doi/10.5555/2627435.2670313}
  \end{APACrefDOI}
\PrintBackRefs{\CurrentBib}

\bibitem [\protect \citeauthoryear {%
Stockwell%
, Kirchner%
, Kuhn%
\BCBL {}\ \BBA {} Seefeld%
}{%
Stockwell%
\ \protect \BOthers {.}}{%
{\protect \APACyear {1997}}%
}]{%
stockwell1997RACM}
\APACinsertmetastar {%
stockwell1997RACM}%
\begin{APACrefauthors}%
Stockwell, W\BPBI R.%
, Kirchner, F.%
, Kuhn, M.%
\BCBL {}\ \BBA {} Seefeld, S.%
\end{APACrefauthors}%
\unskip\
\newblock
\APACrefYearMonthDay{1997}{}{}.
\newblock
{\BBOQ}\APACrefatitle {A new mechanism for regional atmospheric chemistry
  modeling} {A new mechanism for regional atmospheric chemistry
  modeling}.{\BBCQ}
\newblock
\APACjournalVolNumPages{{Journal of Geophysical Research:
  Atmospheres}}{102}{D22}{25847-25879}.
\newblock
\begin{APACrefDOI} \doi{https://doi.org/10.1029/97JD00849} \end{APACrefDOI}
\PrintBackRefs{\CurrentBib}

\bibitem [\protect \citeauthoryear {%
Sturm%
\ \BBA {} Wexler%
}{%
Sturm%
\ \BBA {} Wexler%
}{%
{\protect \APACyear {2020}}%
}]{%
sturm2020mb}
\APACinsertmetastar {%
sturm2020mb}%
\begin{APACrefauthors}%
Sturm, P\BPBI O.%
\BCBT {}\ \BBA {} Wexler, A\BPBI S.%
\end{APACrefauthors}%
\unskip\
\newblock
\APACrefYearMonthDay{2020}{}{}.
\newblock
{\BBOQ}\APACrefatitle {A mass- and energy-conserving framework for using
  machine learning to speed computations: {A} photochemistry example} {A mass-
  and energy-conserving framework for using machine learning to speed
  computations: {A} photochemistry example}.{\BBCQ}
\newblock
\APACjournalVolNumPages{{Geoscientific Model Development}}{13}{9}{4435--4442}.
\newblock
\begin{APACrefDOI} \doi{https://doi.org/10.5194/gmd-13-4435-2020}
  \end{APACrefDOI}
\PrintBackRefs{\CurrentBib}

\bibitem [\protect \citeauthoryear {%
Sturm%
\ \BBA {} Wexler%
}{%
Sturm%
\ \BBA {} Wexler%
}{%
{\protect \APACyear {2022}}%
}]{%
sturm2022conservation}
\APACinsertmetastar {%
sturm2022conservation}%
\begin{APACrefauthors}%
Sturm, P\BPBI O.%
\BCBT {}\ \BBA {} Wexler, A\BPBI S.%
\end{APACrefauthors}%
\unskip\
\newblock
\APACrefYearMonthDay{2022}{}{}.
\newblock
{\BBOQ}\APACrefatitle {Conservation laws in a neural network architecture:
  {E}nforcing the atom balance of a {J}ulia-based photochemical model (v0.2.0)}
  {Conservation laws in a neural network architecture: {E}nforcing the atom
  balance of a {J}ulia-based photochemical model (v0.2.0)}.{\BBCQ}
\newblock
\APACjournalVolNumPages{Geoscientific Model Development}{15}{8}{3417--3431}.
\newblock
\begin{APACrefDOI} \doi{https://doi.org/10.5194/gmd-15-3417-2022}
  \end{APACrefDOI}
\PrintBackRefs{\CurrentBib}

\bibitem [\protect \citeauthoryear {%
Tsitouras%
}{%
Tsitouras%
}{%
{\protect \APACyear {2011}}%
}]{%
tsitouras2011runge}
\APACinsertmetastar {%
tsitouras2011runge}%
\begin{APACrefauthors}%
Tsitouras, C.%
\end{APACrefauthors}%
\unskip\
\newblock
\APACrefYearMonthDay{2011}{}{}.
\newblock
{\BBOQ}\APACrefatitle {Runge–Kutta pairs of order 5(4) satisfying only the
  first column simplifying assumption} {Runge–kutta pairs of order 5(4)
  satisfying only the first column simplifying assumption}.{\BBCQ}
\newblock
\APACjournalVolNumPages{Computers \& Mathematics with
  Applications}{62}{2}{770-775}.
\newblock
\begin{APACrefDOI} \doi{https://doi.org/10.1016/j.camwa.2011.06.002}
  \end{APACrefDOI}
\PrintBackRefs{\CurrentBib}

\bibitem [\protect \citeauthoryear {%
Verwer%
\ \BBA {} Simpson%
}{%
Verwer%
\ \BBA {} Simpson%
}{%
{\protect \APACyear {1995}}%
}]{%
verwer1995AQMstiffness}
\APACinsertmetastar {%
verwer1995AQMstiffness}%
\begin{APACrefauthors}%
Verwer, J.%
\BCBT {}\ \BBA {} Simpson, D.%
\end{APACrefauthors}%
\unskip\
\newblock
\APACrefYearMonthDay{1995}{}{}.
\newblock
{\BBOQ}\APACrefatitle {Explicit methods for stiff ODEs from atmospheric
  chemistry} {Explicit methods for stiff odes from atmospheric
  chemistry}.{\BBCQ}
\newblock
\APACjournalVolNumPages{{Applied Numerical Mathematics}}{18}{1}{413-430}.
\newblock
\begin{APACrefDOI} \doi{https://doi.org/10.1016/0168-9274(95)00068-6}
  \end{APACrefDOI}
\PrintBackRefs{\CurrentBib}

\bibitem [\protect \citeauthoryear {%
Viotti%
, Liuti%
\BCBL {}\ \BBA {} {Di Genova}%
}{%
Viotti%
\ \protect \BOthers {.}}{%
{\protect \APACyear {2002}}%
}]{%
viotti2002nnblackbox}
\APACinsertmetastar {%
viotti2002nnblackbox}%
\begin{APACrefauthors}%
Viotti, P.%
, Liuti, G.%
\BCBL {}\ \BBA {} {Di Genova}, P.%
\end{APACrefauthors}%
\unskip\
\newblock
\APACrefYearMonthDay{2002}{}{}.
\newblock
{\BBOQ}\APACrefatitle {Atmospheric urban pollution: {A}pplications of an
  artificial neural network ({ANN}) to the city of Perugia} {Atmospheric urban
  pollution: {A}pplications of an artificial neural network ({ANN}) to the city
  of perugia}.{\BBCQ}
\newblock
\APACjournalVolNumPages{{Ecological Modelling}}{148}{1}{27-46}.
\newblock
\begin{APACrefDOI} \doi{https://doi.org/10.1016/S0304-3800(01)00434-3}
  \end{APACrefDOI}
\PrintBackRefs{\CurrentBib}

\bibitem [\protect \citeauthoryear {%
Vong%
, Hendrickson%
, Navarro%
\BCBL {}\ \BBA {} Perfors%
}{%
Vong%
\ \protect \BOthers {.}}{%
{\protect \APACyear {2019}}%
}]{%
vong2019additional}
\APACinsertmetastar {%
vong2019additional}%
\begin{APACrefauthors}%
Vong, W\BPBI K.%
, Hendrickson, A\BPBI T.%
, Navarro, D\BPBI J.%
\BCBL {}\ \BBA {} Perfors, A.%
\end{APACrefauthors}%
\unskip\
\newblock
\APACrefYearMonthDay{2019}{}{}.
\newblock
{\BBOQ}\APACrefatitle {Do Additional Features Help or Hurt Category Learning?
  {T}he Curse of Dimensionality in Human Learners} {Do additional features help
  or hurt category learning? {T}he curse of dimensionality in human
  learners}.{\BBCQ}
\newblock
\APACjournalVolNumPages{{Cognitive Science}}{43}{3}{e12724}.
\newblock
\begin{APACrefDOI} \doi{https://doi.org/10.1111/cogs.12724} \end{APACrefDOI}
\PrintBackRefs{\CurrentBib}

\bibitem [\protect \citeauthoryear {%
Wang%
, Estrada%
, Arruda%
\BCBL {}\ \BBA {} Garikipati%
}{%
Wang%
\ \protect \BOthers {.}}{%
{\protect \APACyear {2021}}%
}]{%
wang2021inference}
\APACinsertmetastar {%
wang2021inference}%
\begin{APACrefauthors}%
Wang, Z.%
, Estrada, J.%
, Arruda, E.%
\BCBL {}\ \BBA {} Garikipati, K.%
\end{APACrefauthors}%
\unskip\
\newblock
\APACrefYearMonthDay{2021}{}{}.
\newblock
{\BBOQ}\APACrefatitle {Inference of deformation mechanisms and constitutive
  response of soft material surrogates of biological tissue by full-field
  characterization and data-driven variational system identification}
  {Inference of deformation mechanisms and constitutive response of soft
  material surrogates of biological tissue by full-field characterization and
  data-driven variational system identification}.{\BBCQ}
\newblock
\APACjournalVolNumPages{Journal of the Mechanics and Physics of
  Solids}{153}{}{104474}.
\newblock
\begin{APACrefDOI} \doi{https://doi.org/10.1016/j.jmps.2021.104474}
  \end{APACrefDOI}
\PrintBackRefs{\CurrentBib}

\bibitem [\protect \citeauthoryear {%
Weschler%
}{%
Weschler%
}{%
{\protect \APACyear {2006}}%
}]{%
weschler2006ozone}
\APACinsertmetastar {%
weschler2006ozone}%
\begin{APACrefauthors}%
Weschler, C\BPBI J.%
\end{APACrefauthors}%
\unskip\
\newblock
\APACrefYearMonthDay{2006}{}{}.
\newblock
{\BBOQ}\APACrefatitle {Ozone’s impact on public health: {C}ontributions from
  indoor exposures to ozone and products of ozone-initiated chemistry}
  {Ozone’s impact on public health: {C}ontributions from indoor exposures to
  ozone and products of ozone-initiated chemistry}.{\BBCQ}
\newblock
\APACjournalVolNumPages{{Environmental Health
  Perspectives}}{114}{10}{1489--1496}.
\newblock
\begin{APACrefDOI} \doi{https://doi.org/10.1289/ehp.9256} \end{APACrefDOI}
\PrintBackRefs{\CurrentBib}

\bibitem [\protect \citeauthoryear {%
Yarwood%
\ \protect \BOthers {.}}{%
Yarwood%
\ \protect \BOthers {.}}{%
{\protect \APACyear {2010}}%
}]{%
yarwood2010cb6}
\APACinsertmetastar {%
yarwood2010cb6}%
\begin{APACrefauthors}%
Yarwood, G.%
, Jung, J.%
, Whitten, G\BPBI Z.%
, Heo, G.%
, Mellberg, J.%
\BCBL {}\ \BBA {} Estes, M.%
\end{APACrefauthors}%
\unskip\
\newblock
\APACrefYearMonthDay{2010}{}{}.
\newblock
{\BBOQ}\APACrefatitle {Updates to the {C}arbon {B}ond mechanism for version 6
  ({CB}6)} {Updates to the {C}arbon {B}ond mechanism for version 6
  ({CB}6)}.{\BBCQ}
\newblock
\BIn{} \APACrefbtitle {Paper presented at 9th Annual CMAS Conference} {Paper
  presented at 9th annual cmas conference}\ (\BPGS\ 11--13).
\newblock
\APACaddressPublisher{Chapel Hill, NC}{}.
\PrintBackRefs{\CurrentBib}

\bibitem [\protect \citeauthoryear {%
Zaveri%
\ \BBA {} Peters%
}{%
Zaveri%
\ \BBA {} Peters%
}{%
{\protect \APACyear {1999}}%
}]{%
zaveri1999cbmz}
\APACinsertmetastar {%
zaveri1999cbmz}%
\begin{APACrefauthors}%
Zaveri, R\BPBI A.%
\BCBT {}\ \BBA {} Peters, L\BPBI K.%
\end{APACrefauthors}%
\unskip\
\newblock
\APACrefYearMonthDay{1999}{}{}.
\newblock
{\BBOQ}\APACrefatitle {A new lumped structure photochemical mechanism for
  large-scale applications} {A new lumped structure photochemical mechanism for
  large-scale applications}.{\BBCQ}
\newblock
\APACjournalVolNumPages{{Journal of Geophysical Research:
  Atmospheres}}{104}{D23}{30387-30415}.
\newblock
\begin{APACrefDOI} \doi{https://doi.org/10.1029/1999JD900876} \end{APACrefDOI}
\PrintBackRefs{\CurrentBib}

\bibitem [\protect \citeauthoryear {%
Zhang%
, {Eddy Patuwo}%
\BCBL {}\ \BBA {} {Y. Hu}%
}{%
Zhang%
\ \protect \BOthers {.}}{%
{\protect \APACyear {1998}}%
}]{%
zhang1998review}
\APACinsertmetastar {%
zhang1998review}%
\begin{APACrefauthors}%
Zhang, G.%
, {Eddy Patuwo}, B.%
\BCBL {}\ \BBA {} {Y. Hu}, M.%
\end{APACrefauthors}%
\unskip\
\newblock
\APACrefYearMonthDay{1998}{}{}.
\newblock
{\BBOQ}\APACrefatitle {Forecasting with artificial neural networks: {T}he state
  of the art} {Forecasting with artificial neural networks: {T}he state of the
  art}.{\BBCQ}
\newblock
\APACjournalVolNumPages{{International Journal of Forecasting}}{14}{1}{35-62}.
\newblock
\begin{APACrefDOI} \doi{https://doi.org/10.1016/S0169-2070(97)00044-7}
  \end{APACrefDOI}
\PrintBackRefs{\CurrentBib}

\end{thebibliography}

%
%
%
%
%

\end{document}


%
%


\title{Supporting Information for "Atmospheric chemistry surrogate modeling with sparse identification of nonlinear dynamics"}
%
%

%
%



\authors{Xiaokai Yang\affil{1}, Lin Guo\affil{1}, Zhonghua Zheng\affil{2}, Nicole Riemer\affil{3}, Christopher W. Tessum\affil{1}}


\affiliation{1}{Department of Civil and Environmental Engineering, University of Illinois Urbana-Champaign, Urbana, IL}
\affiliation{2}{Department of Earth and Environmental Sciences, The University of Manchester, Manchester, United Kingdom}
\affiliation{3}{Department of Atmospheric Sciences, University of Illinois Urbana-Champaign, Urbana, IL}

%
%

%

\begin{article}

%
%

\noindent\textbf{Contents of this file}
\begin{enumerate}
\item Figures S1 to S13
\item Tables S1 to S4
\end{enumerate}

\noindent\textbf{Introduction}

 1. Figure S1 to S10 are the SINDy-based surrogate model performance for chemical species prediction on the testing dataset;

 2. Figure S11 to S14 describe the surrogate model computational cost as a function of the number latent species;
 
 3. Tables S1 to S4 contain the reference chemical mechanism, dataset initial conditions, and candidate function library.

\end{article}
\clearpage

 \begin{figure}
\includegraphics[width=\textwidth]{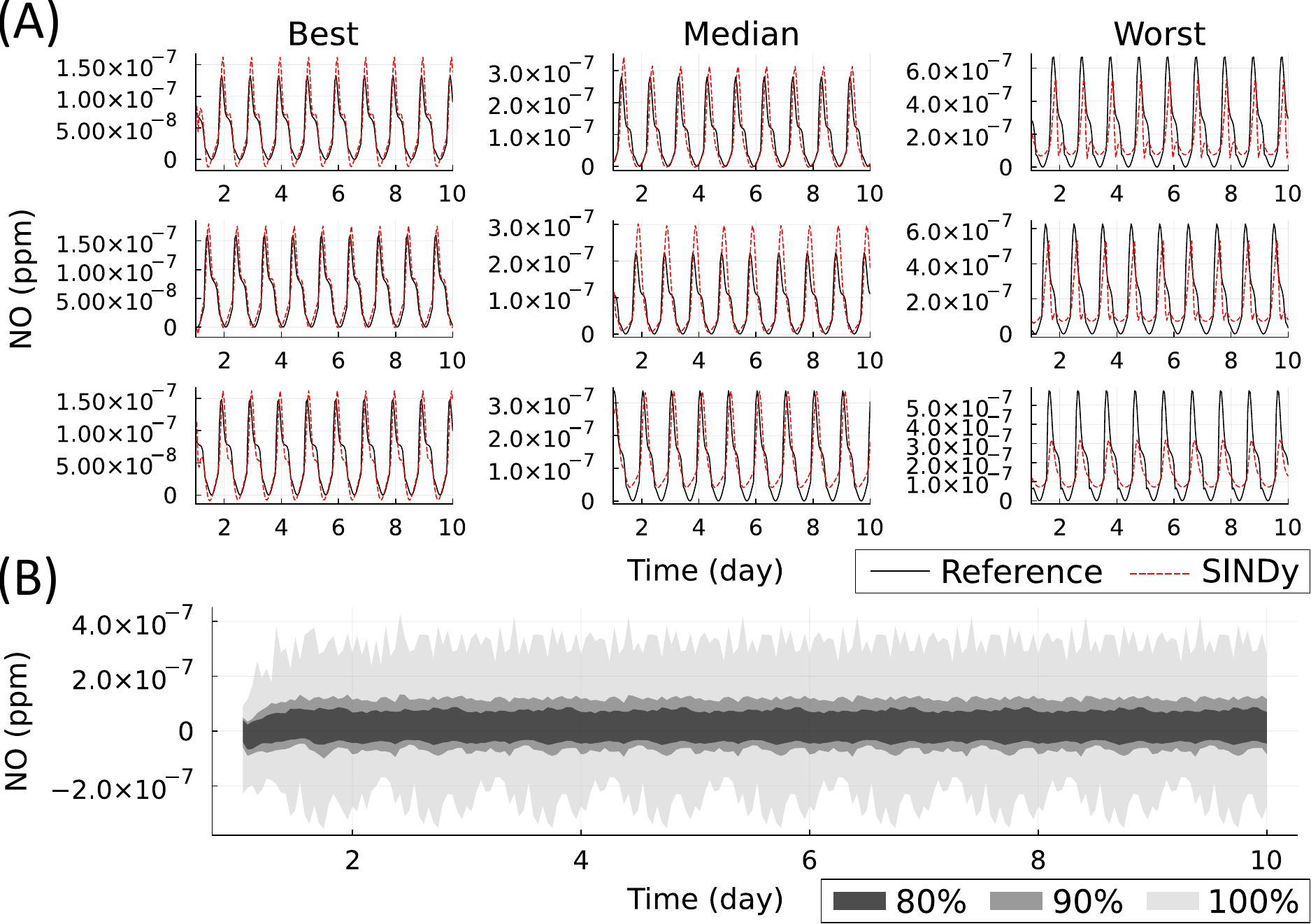}
 \caption{Surrogate model performance of NO prediction. (A) Reference (black solid line) and the surrogate (red dashed line) trajectories for three each of cases with lowest (``best''), median, and highest (``worst'') RSME respectively. (B) Absolute error percentiles for surrogate model testing simulations where the shaded area is the fraction that has a lower absolute error than the value in the legend. }
 \label{fig:figs1}
\end{figure}

 \begin{figure}
\includegraphics[width=\textwidth]{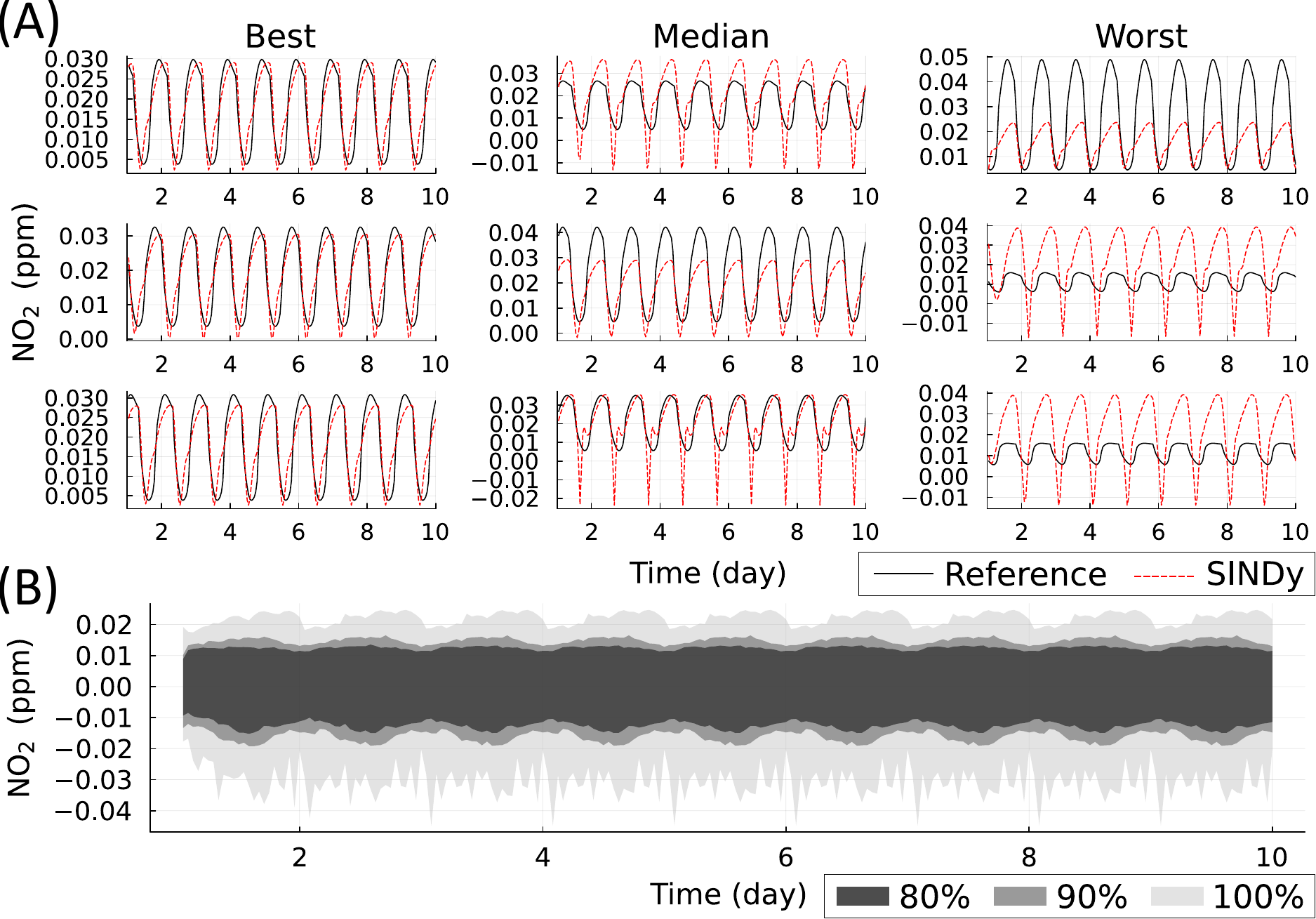}
 \caption{Surrogate model performance of NO\textsubscript{2} prediction. Surrogate model performance. (A) Reference (black solid line) and the surrogate (red dashed line) trajectories for three each of cases with lowest (``best''), median, and highest (``worst'') RSME respectively. (B) Absolute error percentiles for surrogate model testing simulations where the shaded area is the fraction that has a lower absolute error than the value in the legend. }
 \label{fig:figs2}
\end{figure}

 \begin{figure}
\includegraphics[width=\textwidth]{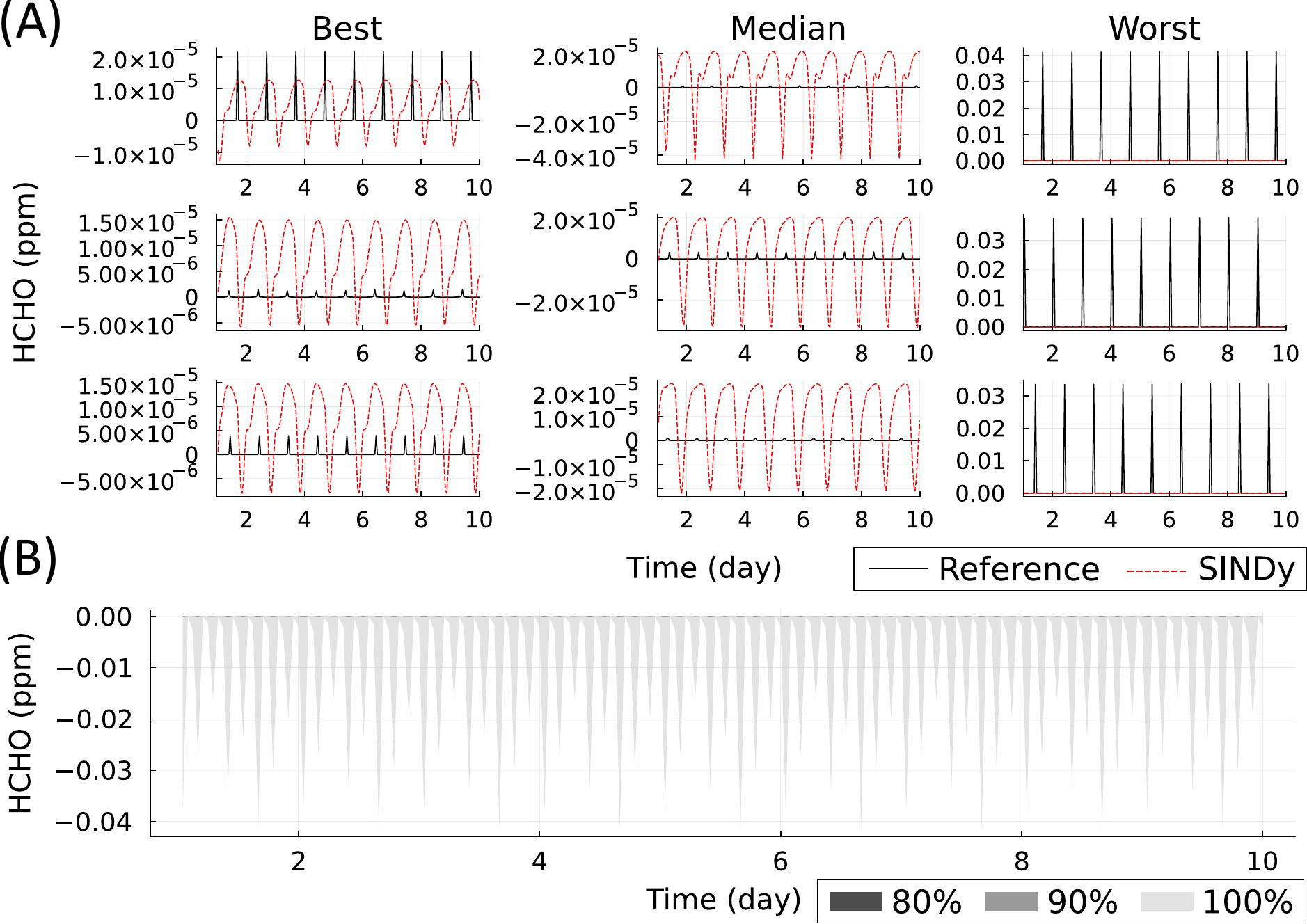}
 \caption{Surrogate model performance of HCHO prediction. (A) Reference (black solid line) and the surrogate (red dashed line) trajectories for three each of cases with lowest (``best''), median, and highest (``worst'') RSME respectively. (B) Absolute error percentiles for surrogate model testing simulations where the shaded area is the fraction that has a lower absolute error than the value in the legend. }
 \label{fig:figs3}
\end{figure}

 \begin{figure}
\includegraphics[width=\textwidth]{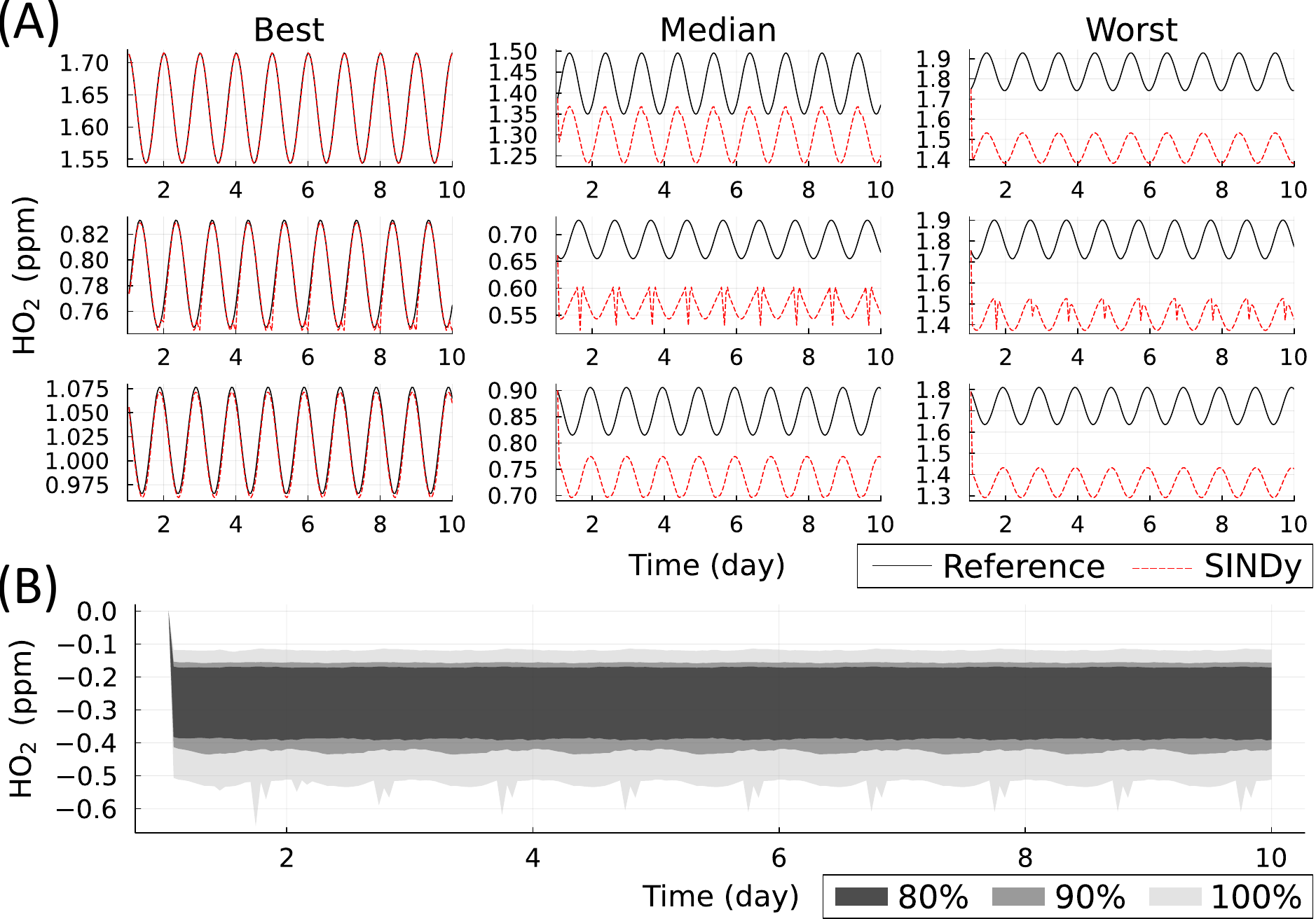}
 \caption{Surrogate model performance of HO\textsubscript{2} prediction. (A) Reference (black solid line) and the surrogate (red dashed line) trajectories for three each of cases with lowest (``best''), median, and highest (``worst'') RSME respectively. (B) Absolute error percentiles for surrogate model testing simulations where the shaded area is the fraction that has a lower absolute error than the value in the legend. }
 \label{fig:figs4}
\end{figure}

 \begin{figure}
\includegraphics[width=\textwidth]{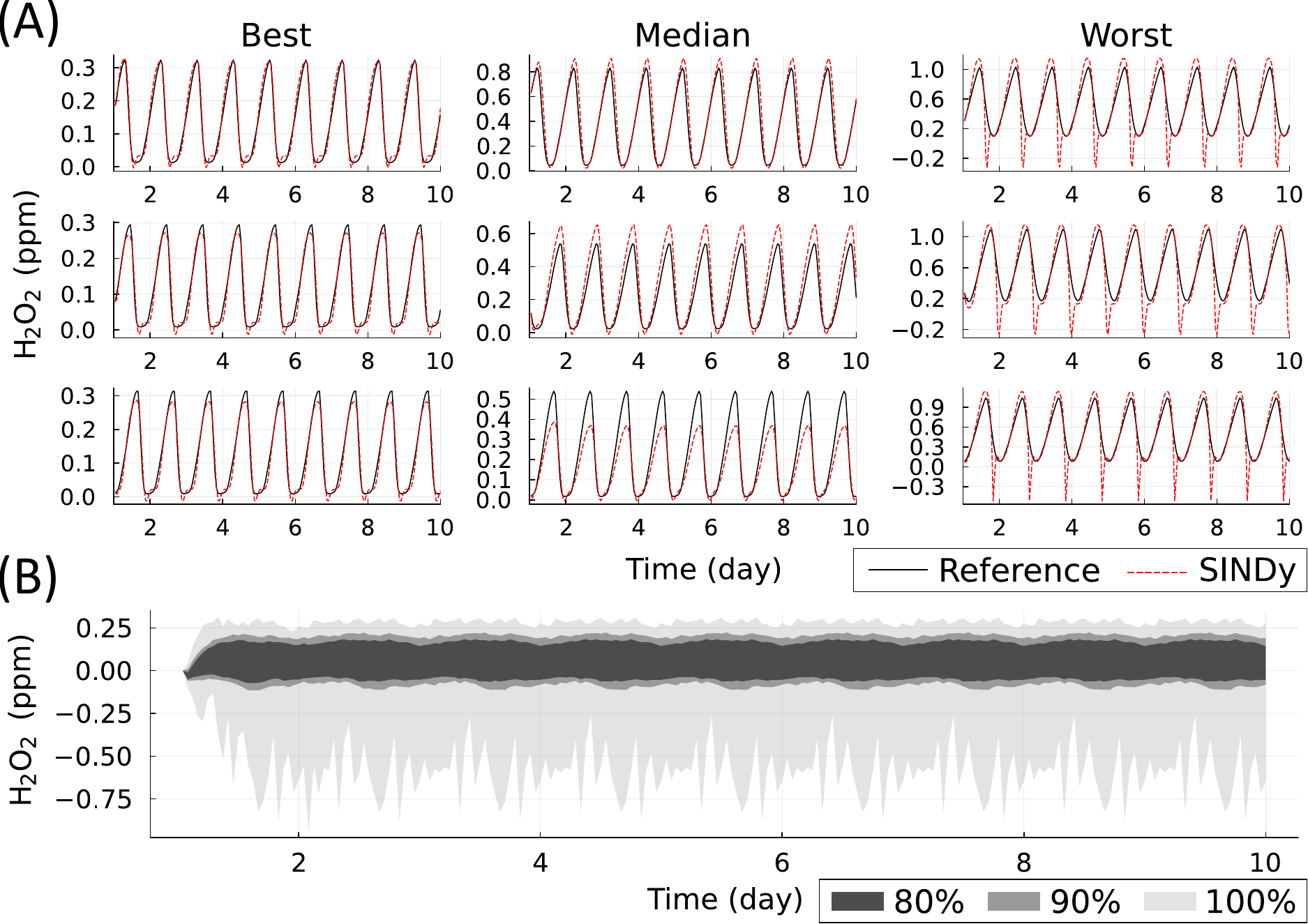}
 \caption{Surrogate model performance of H\textsubscript{2}O\textsubscript{2} prediction. (A) Reference (black solid line) and the surrogate (red dashed line) trajectories for three each of cases with lowest (``best''), median, and highest (``worst'') RSME respectively. (B) Absolute error percentiles for surrogate model testing simulations where the shaded area is the fraction that has a lower absolute error than the value in the legend. }
 \label{fig:figs5}
\end{figure}

 \begin{figure}
\includegraphics[width=\textwidth]{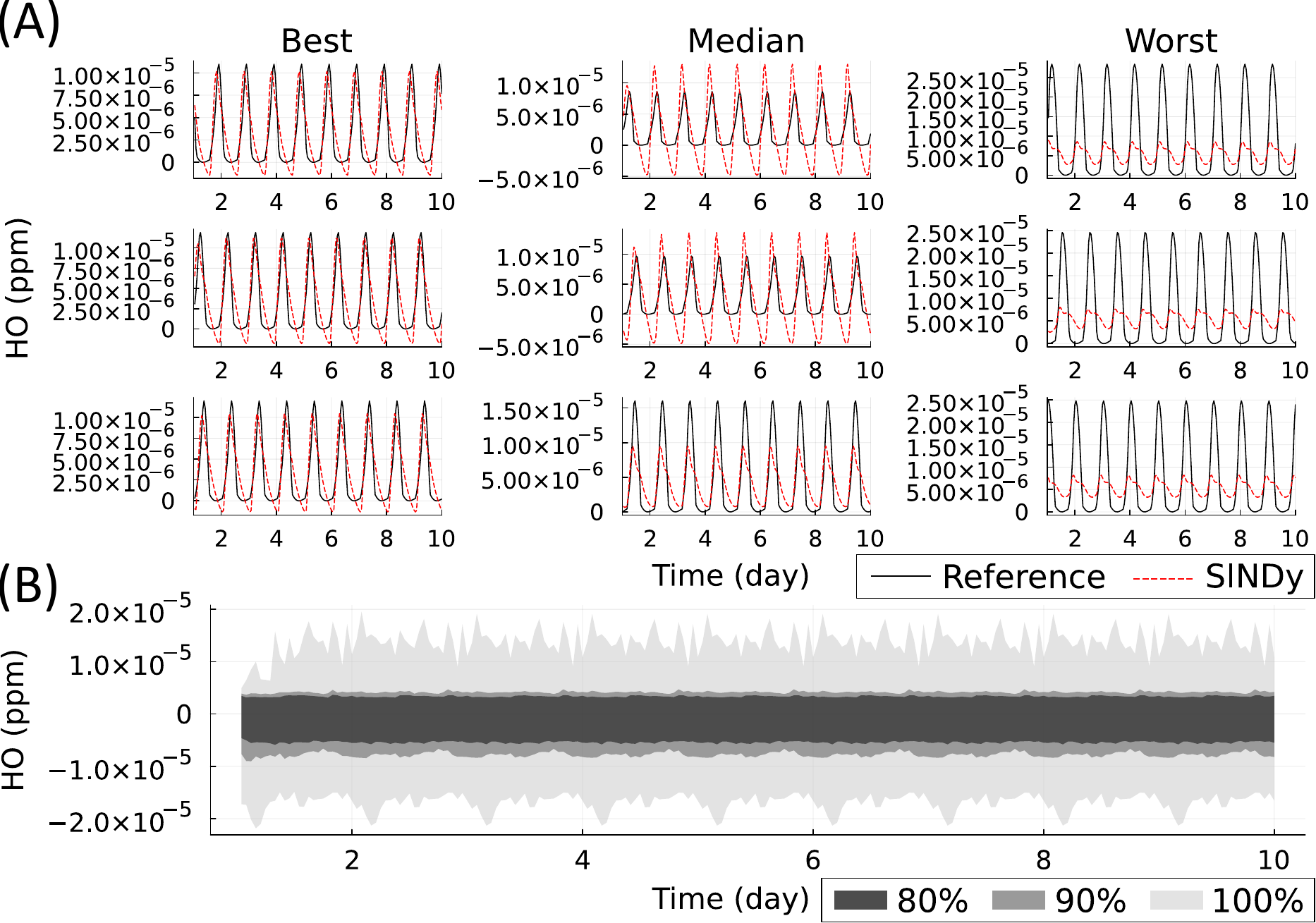}
 \caption{Surrogate model performance of HO prediction. (A) Reference (black solid line) and the surrogate (red dashed line) trajectories for three each of cases with lowest (``best''), median, and highest (``worst'') RSME respectively. (B) Absolute error percentiles for surrogate model testing simulations where the shaded area is the fraction that has a lower absolute error than the value in the legend.}
 \label{fig:figs6}
\end{figure}

 \begin{figure}
\includegraphics[width=\textwidth]{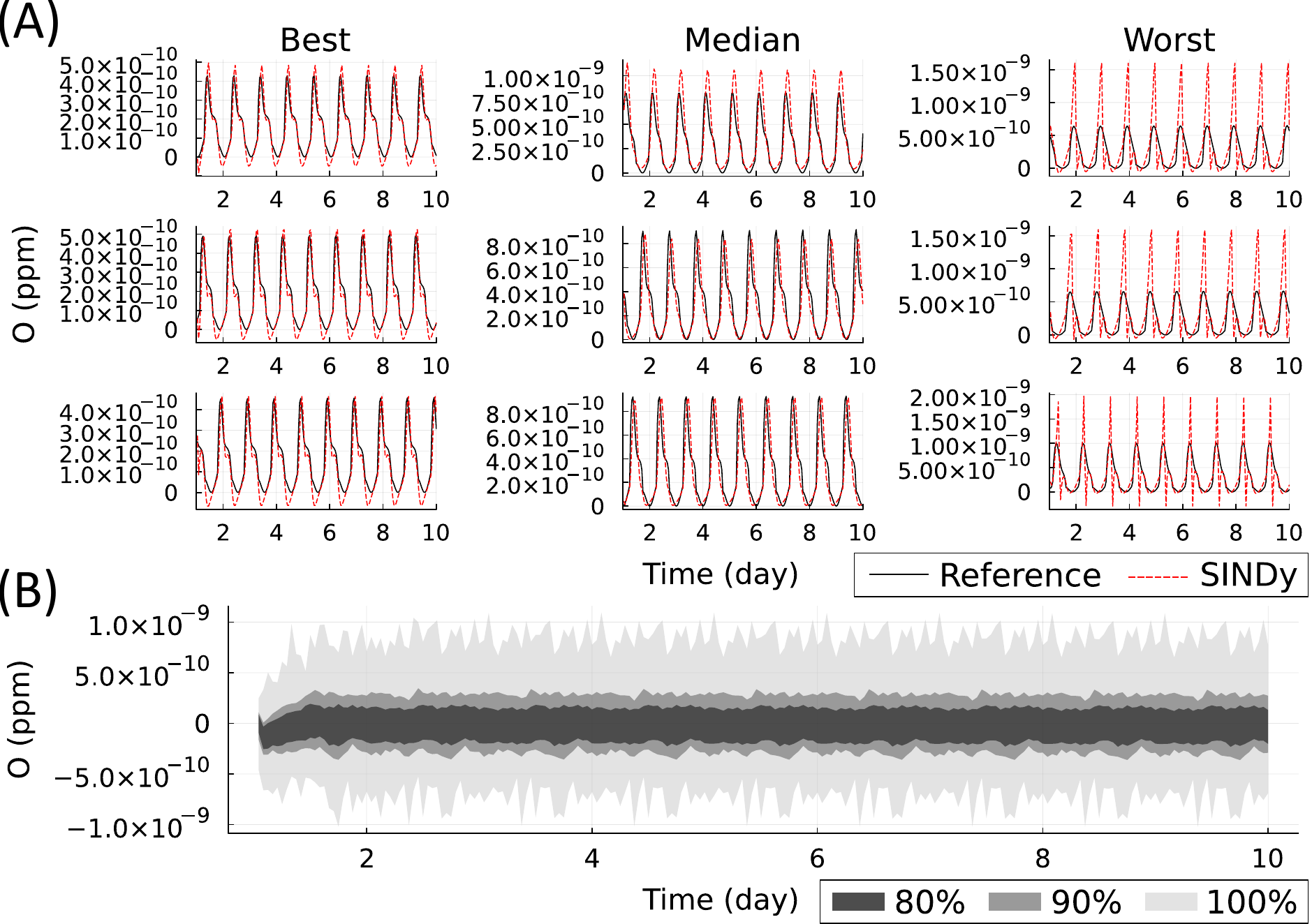}
 \caption{Surrogate model performance of O prediction. (A) Reference (black solid line) and the surrogate (red dashed line) trajectories for three each of cases with lowest (``best''), median, and highest (``worst'') RSME respectively. (B) Absolute error percentiles for surrogate model testing simulations where the shaded area is the fraction that has a lower absolute error than the value in the legend.}
 \label{fig:figs8}
\end{figure}

 \begin{figure}
\includegraphics[width=\textwidth]{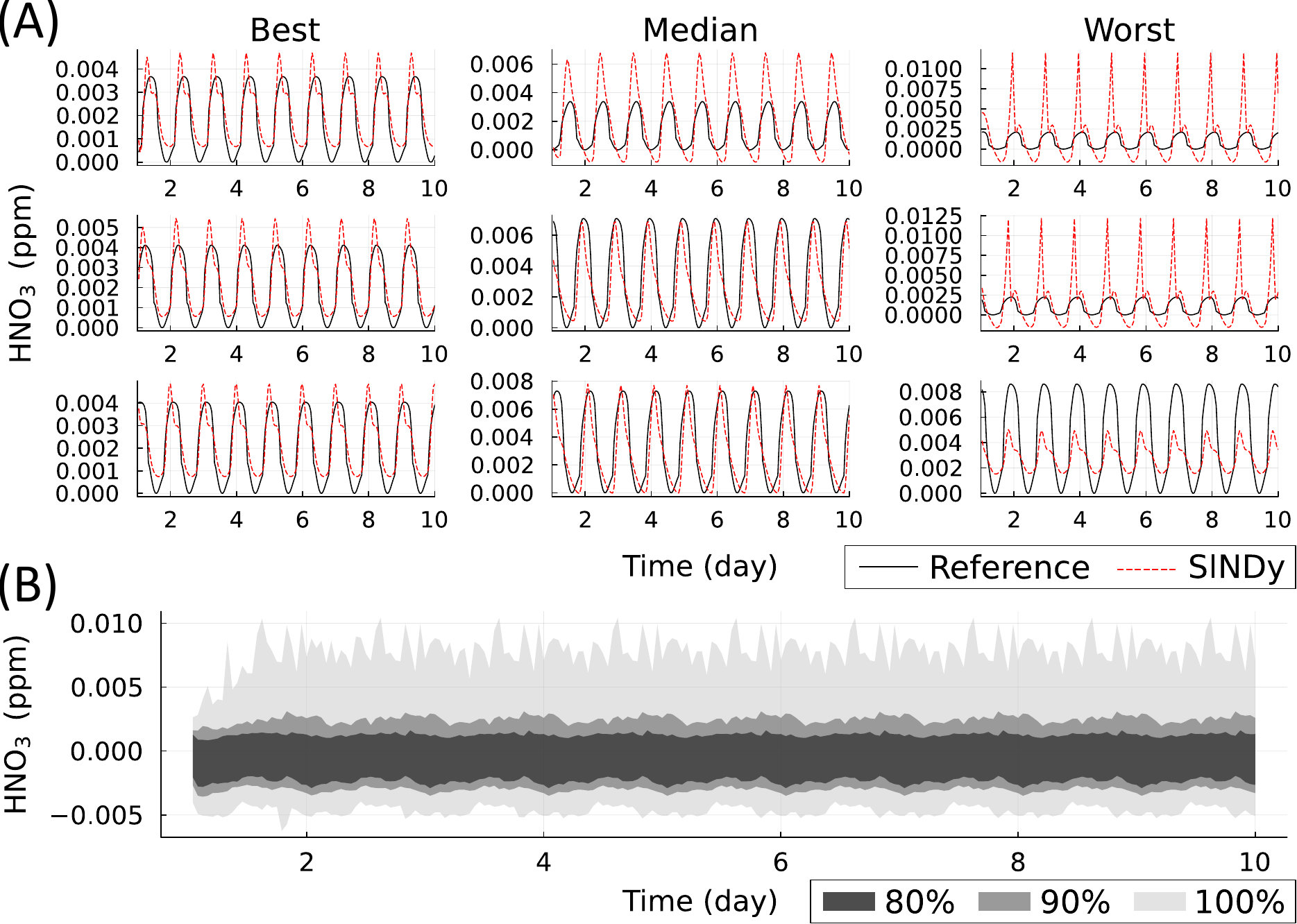}
 \caption{Surrogate model performance of HNO\textsubscript{3} prediction. (A) Reference (black solid line) and the surrogate (red dashed line) trajectories for three each of cases with lowest (``best''), median, and highest (``worst'') RSME respectively. (B) Absolute error percentiles for surrogate model testing simulations where the shaded area is the fraction that has a lower absolute error than the value in the legend.}
 \label{fig:figs7}
\end{figure}

 \begin{figure}
\includegraphics[width=\textwidth]{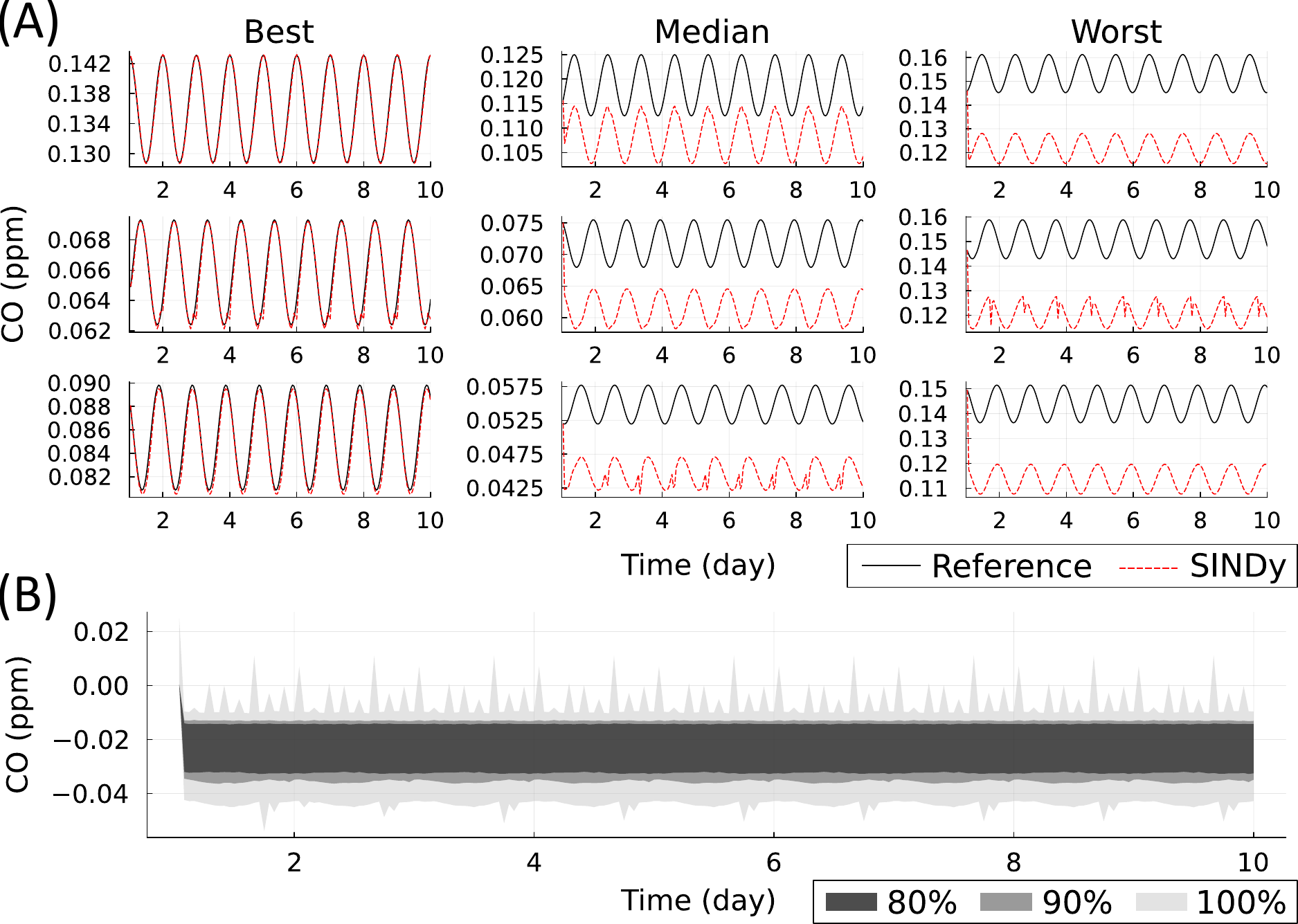}
 \caption{Surrogate model performance of CO prediction. (A) Reference (black solid line) and the surrogate (red dashed line) trajectories for three each of cases with lowest (``best''), median, and highest (``worst'') RSME respectively. (B) Absolute error percentiles for surrogate model testing simulations where the shaded area is the fraction that has a lower absolute error than the value in the legend.}
 \label{fig:figs9}
\end{figure}

 \begin{figure}
\includegraphics[width=\textwidth]{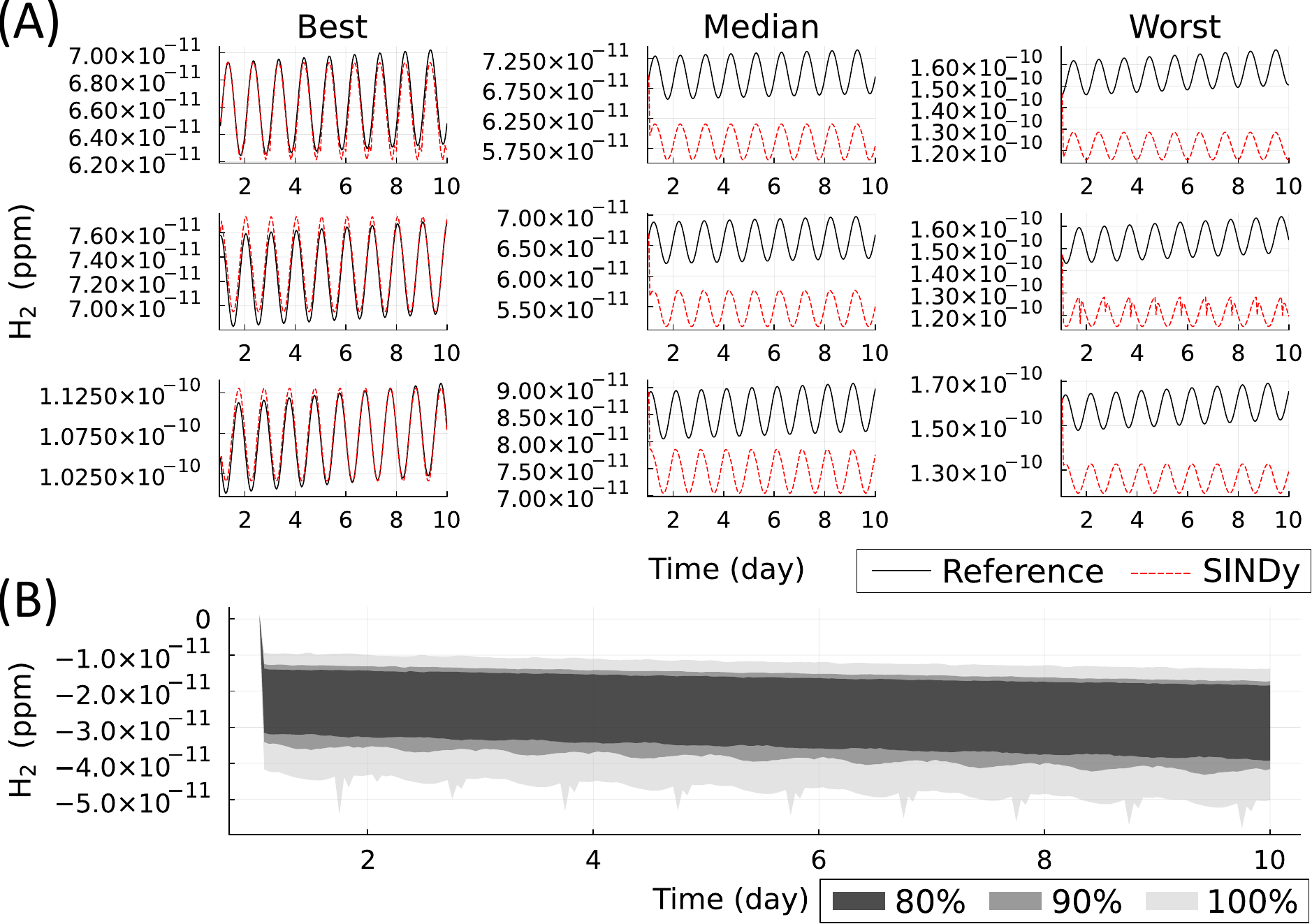}
 \caption{Surrogate model performance of H\textsubscript{2} prediction. (A) Reference (black solid line) and the surrogate (red dashed line) trajectories for three each of cases with lowest (``best''), median, and highest (``worst'') RSME respectively. (B) Absolute error percentiles for surrogate model testing simulations where the shaded area is the fraction that has a lower absolute error than the value in the legend.}
 \label{fig:figs10}
\end{figure}

 \begin{figure}
 \centering
\includegraphics[width=0.65\textwidth]{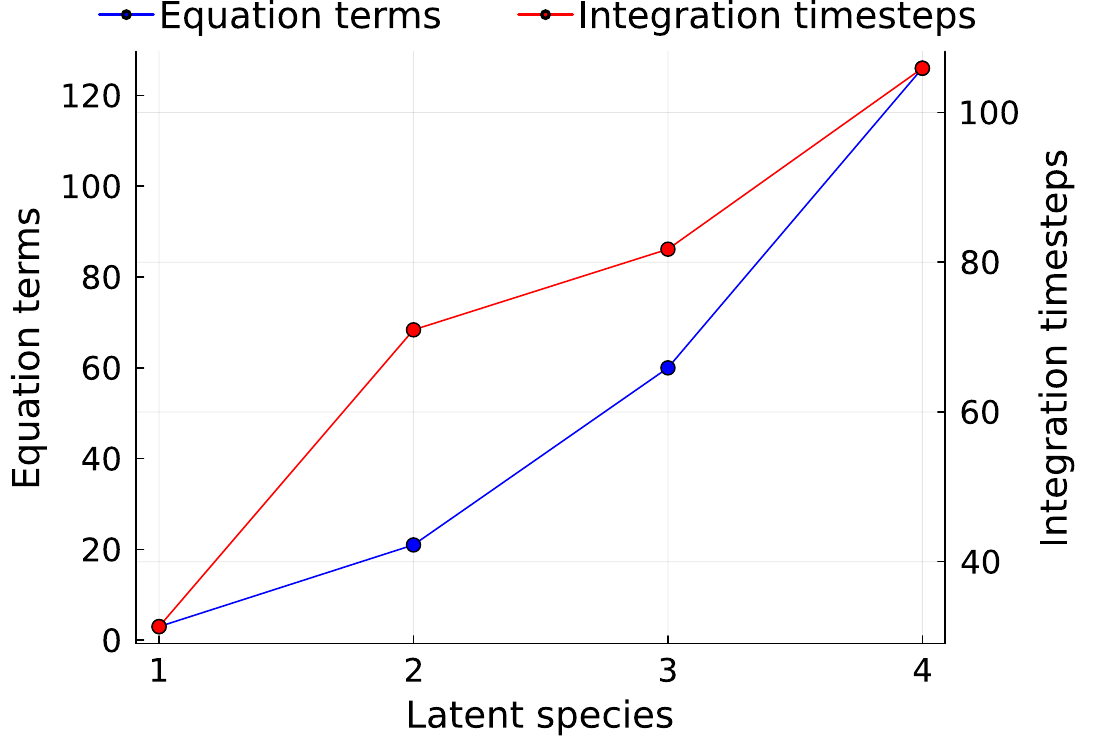}
 \caption{The number of integration timesteps of surrogate model simulation using Tsit5 solver on the testing dataset, and the total number of terms on the RHS of the surrogate ODE, as functions of the number of latent species.}
 \label{fig:figs11}
\end{figure}

 \begin{figure}
 \centering
\includegraphics[width=0.65\textwidth]{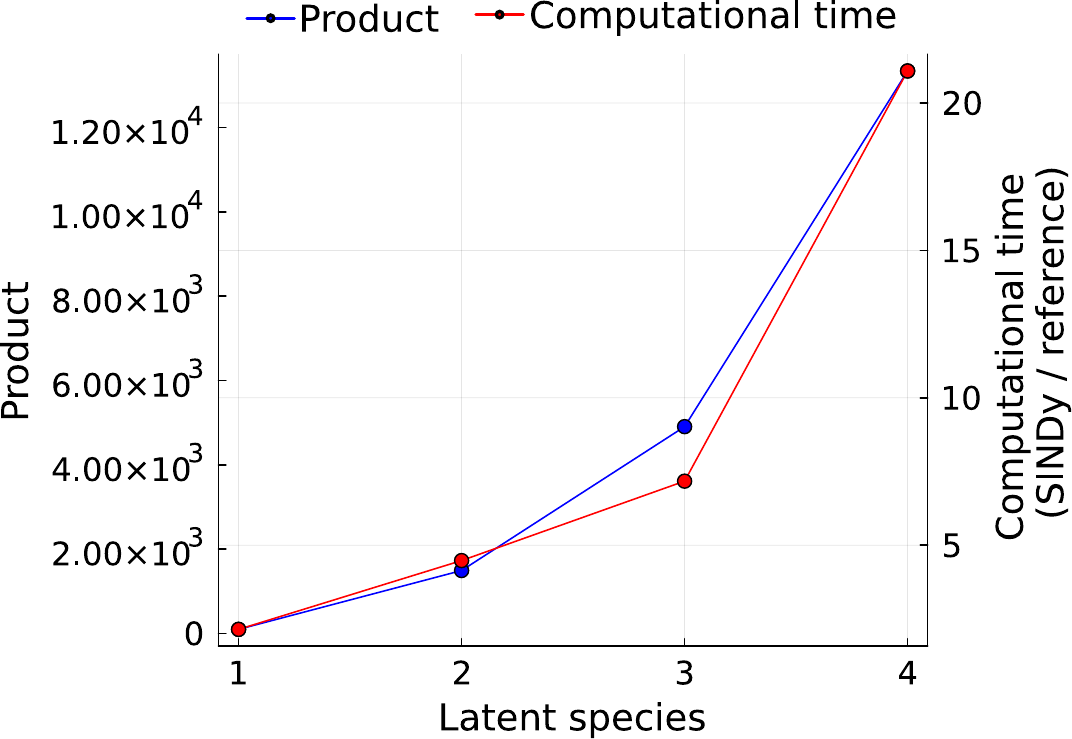}
 \caption{The product of the number of integration timesteps of surrogate model simulation using Tsit5 solver on the testing dataset and the total number of terms on the RHS of the surrogate ODE (the two y-axis value of Figure \ref{fig:figs11}), and the surrogate model computational speed using Tsit5 solver as functions of the number of latent species.}
 \label{fig:figs12}
\end{figure}

 \begin{figure}
 \centering
\includegraphics[width=0.65\textwidth]{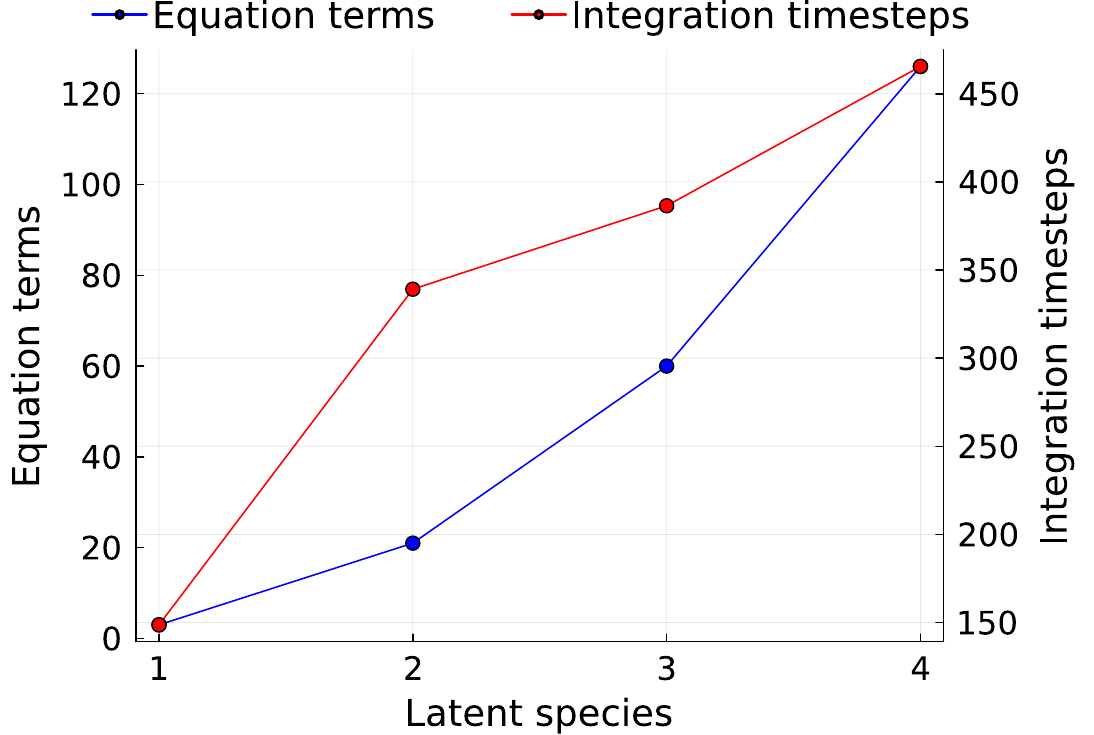}
 \caption{The number of integration timesteps of surrogate model simulation using Rosenbrock23 solver on the testing dataset, and the total number of terms on the RHS of the surrogate ODE, as functions of the number of latent species.}
 \label{fig:figs13}
\end{figure}

 \begin{figure}
 \centering
\includegraphics[width=0.65\textwidth]{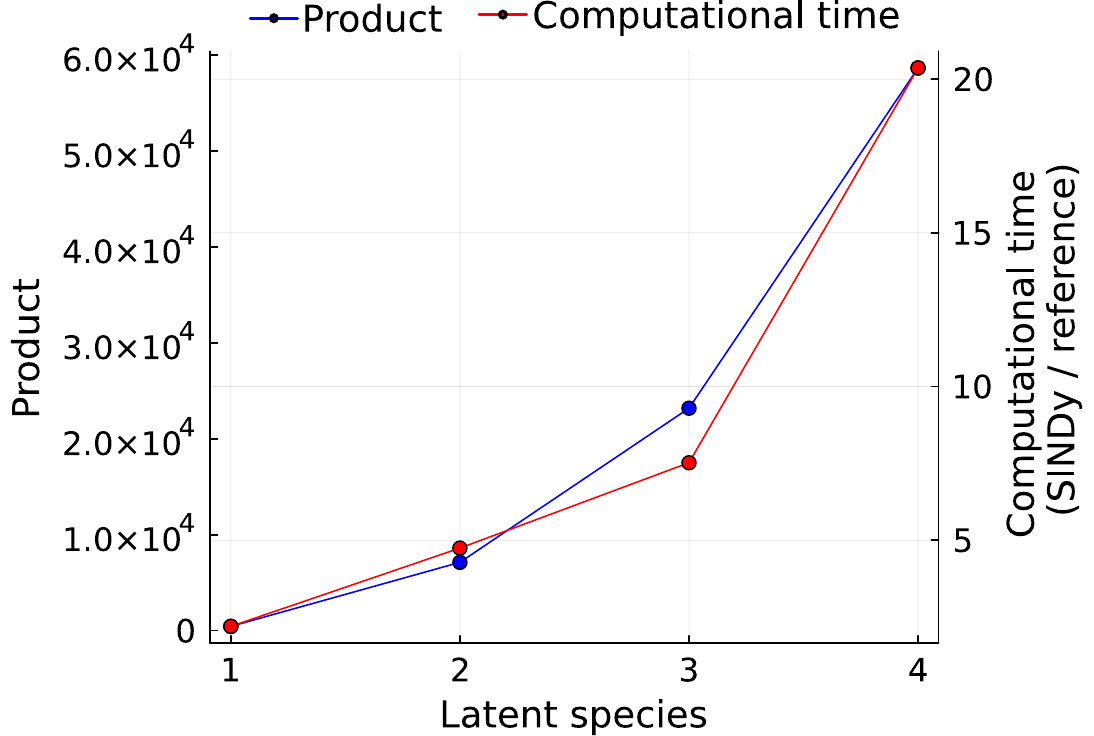}
 \caption{The product of the number of integration timesteps of surrogate model simulation using Rosenbrock23 solver on the testing dataset and the total number of terms on the RHS of the surrogate ODE (the two y-axis value of Figure \ref{fig:figs13}), and the surrogate model computational speed using Rosenbrock23 solver as functions of the number of latent species.}
 \label{fig:figs14}
\end{figure}

\newpage
\begin{table}
\settablenum{S1} 
\caption{Reference chemical mechanism listing}
\centering
\begin{tabular}{c c}
\hline
Number& Reaction                         \\
\hline
(1)    & NO\textsubscript{2} + hv $\rightarrow$ NO + O\\
(2)    & O + O\textsubscript{2}  $\rightarrow$ O\textsubscript{3}\\
(3)    & O\textsubscript{3} + NO  $\rightarrow$ NO\textsubscript{2} + O\textsubscript{2}\\
(4)    & HCHO + hv  $\rightarrow$ 2 HO\textsubscript{2}$\cdot$ + CO\\
(5)    & HCHO + hv  $\rightarrow$ H\textsubscript{2} + CO\\
(6)    & HCHO + HO$\cdot$  $\rightarrow$ HO\textsubscript{2}$\cdot$ + CO + H\textsubscript{2}O\\
(7)    & HO\textsubscript{2}$\cdot$ + NO  $\rightarrow$ OH$\cdot$ + NO\textsubscript{2}\\
(8)    & OH$\cdot$ + NO\textsubscript{2}  $\rightarrow$ HNO\textsubscript{3}\\
(9)    & H\textsubscript{2}O\textsubscript{2} + hv  $\rightarrow$ 2 HO$\cdot$\\
(10)   & H\textsubscript{2}O\textsubscript{2} + HO$\cdot$  $\rightarrow$ H\textsubscript{2}O + HO\textsubscript{2}$\cdot$\\
\hline
\end{tabular}
\end{table}
\newpage

\begin{table}
\settablenum{S2} 

\caption{Initial concentration, emission rates and environmental condition ranges}
\centering
\begin{tabular}{lcccc} 
\hline
Species/Parameters                          & Lower bound & Upper bound & Unit                      & Notes                       \\ 
\hline
O\textsubscript{3}                          & 0           & 0           & ppm                        &     \\ 
OH$\cdot$                                   & 0           & 0           & ppm                        &     \\ 
HNO\textsubscript{3}                        & 0           & 0           & ppm                        &    \\
CO                                          & 0           & 0           & ppm                        &     \\
H\textsubscript{2}                          & 0           & 0           & ppm                        &     \\
O                                           & 0.001       & 0.1         & ppm                        &  (Sturm \& Wexler, 2020, 2022)\\ 
NO                                          & 0.0015      & 0.15        & ppm                        &  (Sturm \& Wexler, 2020, 2022)\\ 
NO\textsubscript{2}                         & 0.0015      & 0.15        & ppm                        &  (Sturm \& Wexler, 2020, 2022)\\ 
HCHO                                        & 0.02        & 2           & ppm                        &  (Sturm \& Wexler, 2020, 2022)\\ 
HO\textsubscript{2}$\cdot$                  & 1           & 10          & ppt                        &  (Sturm \& Wexler, 2020, 2022)\\ 
H\textsubscript{2}O\textsubscript{2}        & 0.001       & 0.01        & ppm                        &  (Sturm \& Wexler, 2020, 2022)\\ 
Pressure                                    & 0.9         & 1.1         & atm                        &   \\ 
Temperature                                 & 288         & 308         & K                          &  \\ 
E (NO\textsubscript{2})                     & 0.0005      & 0.0015      & ppm min\textsuperscript{-1}&  \\ 
E (HCHO)                                    & 0.05        & 0.15        & ppm min\textsuperscript{-1}&  \\ 
E (H\textsubscript{2}O\textsubscript{2})    & 0.0005      & 0.0015      & ppm min\textsuperscript{-1}&  \\ 
$t_0$ (NO\textsubscript{2})                 & 0           & 2$\pi$      & -                          &  \\ 
$t_0$ (HCHO)                                & 0           & 2$\pi$      & -                          &  \\ 
$t_0$ (H\textsubscript{2}O\textsubscript{2})& 0           & 2$\pi$      & -                          & \\ 
Removal rate (O\textsubscript{3})           & 0.02        & 0.02        & ppm min\textsuperscript{-1}&  \\
Removal rate (HNO\textsubscript{3})         & 0.16        & 0.16        & ppm min\textsuperscript{-1}&  \\
Removal rate (HO\textsubscript{2}$\cdot$)   & 0.15        & 0.15        & ppm min\textsuperscript{-1}&  \\
Removal rate (NO\textsubscript{2})          & 0.03        & 0.03        & ppm min\textsuperscript{-1}&  \\
Removal rate (CO)                           & 0.015       & 0.015       & ppm min\textsuperscript{-1}&  \\
Removal rate (H\textsubscript{2})           & 0.06        & 0.06        & ppm min\textsuperscript{-1}&  \\
\hline
\end{tabular}

\end{table}

\begin{table}
\settablenum{S3} 

\caption{Candidate function library and identified coefficient}
\centering
\begin{tabular}{cccc}
\hline
Candidate terms     & $dc_1/dt$    & $dc_2/dt$    & $dc_3/dt$     \\
\hline
$E(\mathrm{NO_2})$  & $0.000140337$  & $0.000530473$  & $-0.000961402$  \\
$E(\mathrm{HCHO})$  & $-0.084512561$ & $0.000193274$  & $-0.034878022$  \\
$E(\mathrm{HO_2H})$ & $-0.000026$    & $-0.001509757$ & $0.000402693$  \\
$c_1$               & $0$            & $0$            & $0$             \\
$c_1^2$             & $0$            & $0$            & $0$             \\
$c_1^3$             & $0.000133205$  & $0$            & $0$             \\
$c_2$               & $0$            & $0$            & $0$             \\
$c_1 c_2$           & $0$            & $0$            & $0$             \\
$c_1^2 c_2$         & $0$            & $0$            & $0$             \\
$c_2^2$             & $0$            & $0$            & $0$             \\
$c_2^2 c_1$         & $0$            & $0.00087524$   & $0.000276296$   \\
$c_2^3$             & $0.004140257$  & $0.000951933$ & $0.015289658$   \\
$c_3$               & $0$            & $0$            & $0$             \\
$c_1 c_3$           & $0$            & $0$            & $0$             \\
$c_1^2 c_3$         & $0$            & $0$            & $0$             \\
$c_2 c_3$           & $0$            & $0$            & $0$             \\
$c_1 c_2 c_3$       & $0$            & $-0.002245274$ & $-0.000552266$  \\
$c_2^2 c_3$         & $-0.013350635$ & $-0.006232409$ & $-0.031021508$  \\
$c_3^2$             & $0$            & $0$            & $0$             \\
$c_3^2 c_1$         & $0$            & $0.001580765$  & $0.000294157$   \\
$c_3^2 c_2$         & $0.014256388$  & $-0.000866642$ & $0.02996446$    \\
$c_3^3$             & $-0.002998676$ & $-0.034416411$ & $0.015746215$   \\
$P T$               & $0.000302426$  & $0$            & $0.000129197$   \\
$P T c_1$           & $-0.000238532$ & $0$           & $-0.0000983$    \\
$P T c_1^2$         & $0$            & $0$            & $0$            \\
$P T c_2$           & $0$            & $0$            & $0$             \\
$P T c_1 c_2$       & $0$            & $0$            & $0$             \\
$P T c_2^2$         & $0$            & $0$            & $0$             \\
$P T c_3$           & $0$            & $0$            & $0$             \\
$P T c_1 c_3$       & $0$            & $0$           & $0$             \\
$P T c_2 c_3$       & $0$            & $0$            & $0.0000684$     \\
$P T c_3^2$         & $0$            & $0$            & $-0.0000579$    \\
$S$                 & $0$            & $0.006076896$  & $-0.006167278$  \\
$S c_1$             & $0$            & $0.00023296$   & $0.000185097$   \\
$S c_1^2$           & $0$            & $0$            & $0$             \\
$S c_2$             & $0.002778896$  & $-0.015569198$ & $0.026215915$   \\
$S c_1 c_2$         & $0$            & $-0.001160901$ & $-0.000125573$  \\
$S c_2^2$           & $-0.004595275$ & $-0.008164743$ & $-0.017004855$  \\
$S c_3$             & $-0.001811391$ & $-0.002188497$ & $-0.013587148$  \\
$S c_1 c_3$         & $0$            & $0.001113011$  & $0.00059286$6   \\
$S c_2 c_3$         & $0.012380522$  & $-0.02824794$  & $0.04146502$    \\
$S c_3^2$           & $-0.005572064$ & $-0.026953527$ & $0.021190034$   \\
\hline
\end{tabular}

\end{table}

\begin{table}
\settablenum{S4} 

\caption{Candidate function terms}
\centering
\begin{tabular}{cccccc}
\hline
$dc_1/dt$          & $\mathrm{Fraction}_1^a$($\%$)& $dc_2/dt$          & $\mathrm{Fraction}_2^a$($\%$)& $dc_3/dt$          & $\mathrm{Fraction}_3^a$($\%$)\\
\hline
$P T$              & 46.60616007   & $S$                & 40.22956099   & $P T$              & 73.125069     \\
$E(\mathrm{HCHO})$ & 41.51107239   & $E(\mathrm{HO_2H})$& 29.83959476   & $P T c_1$          & 17.68190058   \\
$P T c_1$          & 11.67954024   & $E(\mathrm{NO_2})$ & 10.48262055   & $S$                & 3.728102206   \\
$E(\mathrm{NO_2})$ & 0.068947811   & $S c_2$            & 7.591200788   & $S c_2$            & 1.167185825   \\
$S c_2 c_3$        & 0.047684654   & $S c_2 c_3$        & 4.376075433   & $E(\mathrm{HO_2H})$& 1.053083679   \\
$S c_2$            & 0.033686684   & $E(\mathrm{HCHO})$ & 3.818336484   & $E(\mathrm{NO_2})$ & 0.870137828   \\
$S c_3$            & 0.02180811    & $S c_3$            & 1.059766264   & $S c_3$            & 0.600790666   \\
$E(\mathrm{HO_2H})$& 0.012773907   & $S c_3^2$          & 0.954729536   & $S c_2 c_3$        & 0.586557572   \\
$S c_3^2$          & 0.004907073   & $S c_1$            & 0.490002056   & $E(\mathrm{HCHO})$ & 0.560735347   \\
$S c_2^2$          & 0.004102779   & $S c_2^2$          & 0.293202399   & $P T c_2 c_3$      & 0.208451438   \\
$c_3^2 c_2$        & 0.00290523    & $c_3^3$            & 0.28016566    & $P T c_3^2$        & 0.175271222   \\
$c_2^2 c_3$        & 0.002739385   & $S c_1 c_2$        & 0.179842833   & $S c_3^2$          & 0.068537325   \\
$c_1^3$            & 0.00220939    & $S c_1 c_3$        & 0.171244853   & $S c_2^2$          & 0.055760758   \\
$c_2^3$            & 0.000855379   & $c_1 c_2 c_3$      & 0.079937751   & $S c_1$            & 0.035550659   \\
$c_3^3$            & 0.000606905   & $c_3^2 c_1$        & 0.055894596   & $c_2^2 c_3$        & 0.023377776   \\
                   &               & $c_2^2 c_3$        & 0.051435797   & $c_3^2 c_2$        & 0.022426765   \\
                   &               & $c_2^2 c_1$        & 0.031375454   & $c_3^3$            & 0.011704591   \\
                   &               & $c_2^3$            & 0.007910352   & $c_2^3$            & 0.011601608   \\
                   &               & $c_3^2 c_2$        & 0.007103446   & $S c_1 c_3$        & 0.008329242   \\
                   &               &                    &               & $c_1 c_2 c_3$      & 0.001795401   \\
                   &               &                    &               & $S c_1 c_2$        & 0.001776336   \\
                   &               &                    &               & $c_3^2 c_1$        & 0.000949757   \\
                   &               &                    &               & $c_2^2 c_1$        & 0.000904414   \\
\hline
\end{tabular}

\tablenotetext{a}{$\mathrm{Fraction}_i = \frac{\theta_n \xi_n}{\sum_{n=1}^{N}\theta_n \xi_n} \times 100$ where $\theta_n$ is the candidate function term $n$ (substituted by its average value in the testing cases), and $\xi_n$ is the identified coefficient of term $n$, and $N$ is the total number of candidate terms in surrogate equation $i$.}
\end{table}
